\documentclass[english]{IEEEtran}

\usepackage[T1]{fontenc}
\usepackage{xcolor}
\usepackage{amsmath}
\usepackage{amsthm}
\usepackage{amsfonts}
\usepackage{graphicx}
\usepackage{mathtools}
\usepackage{amssymb}
\usepackage{dblfloatfix}

\makeatletter
\theoremstyle{plain}
\newtheorem{thm}{\protect\theoremname}
\theoremstyle{plain}
\newtheorem{lem}[thm]{\protect\lemmaname}
\theoremstyle{plain}

\theoremstyle{definition}
\newtheorem{example}[thm]{\protect\examplename}
\newtheorem{definition}[thm]{\protect\definitionname}

\usepackage[caption=false,font=footnotesize]{subfig}

\makeatother

\providecommand{\corollaryname}{Corollary}
\providecommand{\lemmaname}{Lemma}
\providecommand{\theoremname}{Theorem}
\providecommand{\examplename}{Example}
\providecommand{\propositionname}{Proposition}
\providecommand{\claimname}{Claim}
\providecommand{\conjecturename}{Conjecture}
\providecommand{\definitionname}{Definition}

\DeclareMathOperator{\Tr}{Tr}

\usepackage[textsize=tiny]{todonotes}

\begin{document}
\title{Faithful Simulation of Distributed Quantum Measurement
with Coding for Computing}
\author{Anders H{\o}st-Madsen, \emph{Fellow, IEEE}\thanks{The author is with the Department of Electrical and Computer Engineering, University of Hawaii, Manoa,
 Honolulu, HI, 96822, Email: ahm@hawaii.edu. 
 The research was funded in part by the NSF grant CCF-2324396.
 }}
\maketitle

\begin{abstract}
This paper considers a two-terminal problem in which Alice and Bob aim to perform a joint measurement on a bipartite quantum system $\rho^{AB}$.
Alice transmits the results of her measurements to Bob over a classical channel, and the two share common randomness.
The central question is: what is the minimum amount of communication and common randomness required to faithfully simulate the measurement? This paper derives an achievable rate region.

\end{abstract}


\section{Introduction}
The aim of network information theory \cite{ElGamalKimBook}
is for nodes to obtain certain desired information generated
by sources. In some cases, this is the raw information generated
by the sources, but in many cases it is a function of the information.
For example, a node might only need the sum or average of certain measurements or need to make a decision based on the data..
The required rates for
computing functions in networks are therefore a widely
studied problem in classical information theory, for example
\cite{OrlitskyRoche01, DoshiMedardEffrosAl10,FeiziMedard14,GuangYeungAl19,Malak22, Gaspar08,NazerGastpar08,NazerGastpar07, SoundararajanVishwanathAl12, AppuswamyZegerAl11, LiAl23, LiMaddah-AliAl17, ZhuGunduz21, BergerZhang94, ViswanathanBerger97, HeAl16}
.
Computing sums or linear functions in particular has found many solutions \cite{Gaspar08,NazerGastpar08,NazerGastpar07, SoundararajanVishwanathAl12, AppuswamyZegerAl11, LiMaddah-AliAl17, BergerZhang94, ViswanathanBerger97, HeAl16}, which has also been generalized to the quantum setting in
a number of papers, for example \cite{SohailPradhan22,YaoJafar24,AllaixJafarAl25}.
Finding rate regions for general functions has fewer
solutions. Orlitsky and Roche \cite{OrlitskyRoche01} found
the exact rate required for the following setup: Alice knows
a random variable $U$ and Bob knows $V$ and the goal is
for Bob to calculate a function $g(U,V)$; how much
information does Alice need to transmit? Generalizations
of this problem have been considered in a number of papers \cite{ElGamalKimBook,DoshiMedardEffrosAl10,FeiziMedard14,GuangYeungAl19,Malak22}, but
the only case where a general and complete solution seems to be known is the one
considered by Orlitsky and Roche.

Measurements are, of course, central to quantum theory.
A version of the networked function computation problem for
quantum measurements is as follows. 
Nodes $A,B,C\ldots$ have a multipartite system represented
by a density operator $\rho^{ABC\ldots}$. One node, say A,
wants to perform a global measurement represented
by POVM $\{\Lambda^{ABC\ldots}_a\}_a$, but has only its
local quantum system. Consequently, the measurement must be executed using local quantum instruments at each node, with the outcomes transmitted to a destination node. How much transmission is needed
to find $\{\Lambda^{ABC\ldots}_a\}_a$, which is, of course, classical information? An application could be a distributed quantum computer,
where one would like to extract a classical result that depends on
all the quantum states.

The question is what it means to find $\{\Lambda^{ABC\ldots}_a\}_a$.
Winter \cite{Winter2004} divides the outcome from
a measurement into "meaningful" (intrinsic) and "not meaningful" (extrinsic) information. In
terms of communications, one could say that only
intrinsic information needs to be communicated. On the
other hand, no distortion of $\{\Lambda^{ABC\ldots}_a\}$
is allowed. This leads to the idea of
faithful simulation of measurements  
\cite{Winter2004}. The idea is as follows. Suppose that
in a two node system
Alice performs some measurements and want to
transmit the result to Bob.
Alice performs a measurement on a quantum state $\rho$
and sends some classical bits to Bob, who intends
to \emph{faithfully} recover Alice's measurement,
preserving correlation with the reference system.
The key observation is that if Alice and Bob have
common randomness, the number of bits transmitted
from Alice to Bob can be decreased below that of
classical data compression of Alice's measurement 
outcomes, 
while still preserving the correlation with
the reference system. We refer the reader
to the overview paper \cite{WildeAl2012} and the
paper \cite{AtifHeidariSandeep22} for more details
about faithful simulation and applications of it. 
One could think of faithful
simulation as a method to concretely reduce
communication rates, but it can also be used to
solve many other problems in quantum information
theory such as rate distortion and local purity distilation \cite{WildeAl2012,AtifHeidariSandeep22}. This
paper is only concerned with finding communication
rates leaving possible applications to later papers.

In \cite{AtifHeidariSandeep22}
the authors considered the following problem. Alice and Bob have a shared
bipartite quantum system $\rho^{AB}$. Alice makes some
measurements with a POVM $\{\Lambda^A_u\}$ on $\rho^A$ and
Bob measures with $\{\Lambda^B_v\}$ on $\rho^B$. Alice, Bob, and Charlie
share some common randomness, and Alice and Bob transmit some classical
information to Charlie. The goal is essentially for Charlie to
faithfully simulate a function $z=g(u,v)$ of the measurements.

\begin{figure}[hbt]
\begin{centering}
	
  \includegraphics[width=2.5in]{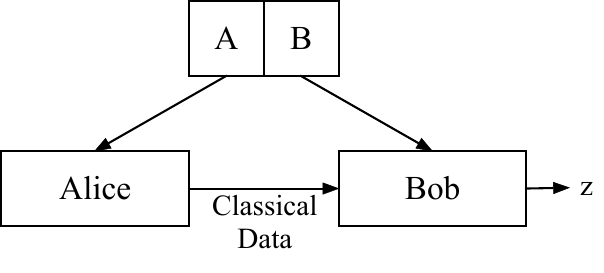}
 \vspace{-0.2in} \caption{System model. The aim if
  for Bob to faithfully simulate $\Lambda_z^{AB}=\sum_{u,v:z=g(u,v)}
  \Lambda_u^A\otimes\Lambda_v^B$ on
  the bipartite system $\rho^{AB}$.\label{fig:system}}
\vspace{-0.1in}
\end{centering}
\end{figure}

In this paper, we consider the seemingly simpler problem in
Fig. \ref{fig:system}.
The difference from the problem considered in \cite{AtifHeidariSandeep22} is that  the three terminal
function computation problem that they consider is not solved in general even
for classical systems. On the other hand,
as was mentioned above,  the solution of the two-terminal problem in Fig. \ref{fig:system} is known, as probably the only network, 
in the classical case; that solution
will guide us to a quantum solution. Another difference is
that Charlie in \cite{AtifHeidariSandeep22} has no quantum information.
On the other hand, in Fig. \ref{fig:system} Bob has access to the quantum
system $\rho^B$ and can use that to decode Alice's transmission in
addition to measuring $\{\Lambda^B_v\}$, similar to faithful simulation
with quantum side information \cite{WildeAl2012,DevetakWinter03,Renes12}.

The paper largely follows the notation and terminology established
in \cite{WildeBook}. We will use theorems and lemmas of \cite{WildeBook}
without repeating them here. We adopt the definition of strong typicality from \cite{WildeBook}. example, the strongly classical typical set is defined as
\begin{align}
  T_\delta^{X^n}&=\left\{
x^n:\forall x\in\mathcal{X}:\left|\frac 1 n N(x|x^n)-p(x)\right|\leq\delta\text{ if p(x)>0 },\right.\nonumber\\
&\left.\frac 1 n N(x|x^n)=0\text{ if p(x)=0 }
  \right\} \label{eq:TypDef}
\end{align}
where $\frac 1 n N(x|x^n)$ is the count of $x$ in $x^n$. 
The paper \cite{OrlitskyRoche01} uses robust typicality (see also \cite{ElGamalKimBook}).
The main feature of robust typicality is that 
$\frac 1 n N(x|x^n)=0$ if p(x)=0, which it shares with
the above definition. We can therefore mostly use the results in
\cite{OrlitskyRoche01}.
We use implicit notation for probabilities when the meaning is unambiguous, e.g.,
$p(u^n)=p^n_U(u^n)$.
A sub-POVM is a set of operators $\{\Lambda_x\}_x$ with
$\Lambda_x\geq 0$ and $\sum_x \Lambda_x\leq I$.
For an integer $a$ $[a]=\{1,2,\ldots a\}$. We use $\|\cdot\|_1$ 
to denote trance norm and trace distance.

\section{Problem Statement}

We consider a bipartite composite quantum system $(A,B)$ represented
by a density operator $\rho^{AB}$ on the Hilbert space
$\mathcal{H}_A\otimes\mathcal{H}_B$. We denote the purification
of $\rho^{AB}$ as $\phi^{RAB}$ for some
reference system $R$. Alice has access to $\rho^A$
and Bob has access to $\rho^B$. The aim is for Bob to perform
the measurement $\{\Lambda_z^{AB}\}_{z\in\mathcal{Z}}$
\begin{align*}
  \Lambda_z^{AB}=\sum_{u\in\mathcal{U},v\in\mathcal{V}:z=g(u,v)}
  \Lambda_u^A\otimes\Lambda_v^B
\end{align*}
where $\{\Lambda_u^A\}_{u\in\mathcal{U}}$ and $\{\Lambda_v^B\}_{v\in\mathcal{V}}$ are POVMs on $\mathcal{H}_A$ respectively
$\mathcal{H}_B$ and $g$ is a deterministic function. 
This is called a separable decomposition with
deterministic integration in \cite{AtifHeidariSandeep22}.
Alice and Bob are given $n$ copies of their states,
$(\rho^{A})^{\otimes n},(\rho^{B})^{\otimes n}$, and
the measurement is performed $n$ times,
\begin{align*}
  \Lambda_{z^n}^{AB}=\sum_{u\in\mathcal{U},v\in\mathcal{V}:z^n=g^n(u^n,v^n)}
  (\Lambda_u^A)^{\otimes n}\otimes(\Lambda_v^B)^{\otimes n}
\end{align*}
where $g^n(u^n,v^n)=(g(u_1,v_1),g(u_2,v_2),\ldots)$.

We only require Bob to faithfully simulate $\{\Lambda_{z^n}^{AB}\}_{z^n\in\mathcal{Z}^n}$ when
$n$ becomes large using classical transmission from Alice
and common randomness with Alice. 
Let $p_M(m)$ be the common randomness distribution.
For each value of $m$ Alice has a sub-POVM
$\{\Gamma^{(m)}_j\}_{j=1}^s$ that jointly measures
on the tensor-power state $(\rho^A)^{\otimes n}$; Alice transmits
the measurement outcome $j$ to Bob. Bob uses some POVM
$\{\Lambda^B_x\}_{x\in\mathcal{X}}$ on the tensor-power state $(\rho^B)^{\otimes n}$
as $(\Lambda^B_x)^{\otimes n}$ ($x=v$ is one possibility, but we will allow other values of $x$).
Bob has a function $f:[s]\times [M]\times \mathcal{X}^n\to \mathcal{Z}^n$ to calculate $z^n$.
This gives an approximate measurement of $z^n$:
\begin{align*}
  \tilde\Lambda_{z^n}^{AB}=\sum_{j,m,x^n:z^n=f(j,m,x^n)}p_M(m)
  \Gamma^{(m)}_j\otimes\Lambda_{x^n}^B
\end{align*}
The requirement is that $\tilde\Lambda_{z^n}^{AB}$ faithfully
simulates $\Lambda_{z^n}^{AB}$ in the following sense \cite{WildeAl2012}: for
all $\epsilon>0$ and sufficiently large $n$, 
\begin{align}
  \sum_{z^n}\|\sqrt{\omega}(\tilde\Lambda_{z^n}^{AB}-\Lambda_{z^n}^{AB})\sqrt{\omega}\|_1\leq\epsilon \label{eq:faithful}
\end{align}
where $\omega=(\rho^{AB})^{\otimes n}$. Arguments for this criterion can be found in \cite{Winter2004,WildeAl2012}.
Notice that there is no reason for Bob to do an approximate measurement,
as we only consider the communications cost from Alice to Bob; an
extension could be to also consider compression of $z^n$. 

The results in \cite{AtifHeidariSandeep22} can be adapted to this system,
and their Theorem 4 gives\footnote{As in \cite{WildeBook} we
use $R$ for both the rate and the reference system.}
\begin{align}
  R &\geq I(U;RB)-I(U;V) \label{eq:AtifR}\\
  R+S &\geq H(U|V) \label{eq:AtifRS}
\end{align}
where $R$ is the rate of communications and $S$ is the rate of common
randomness.

In the classical case, the above problem reduces to the one
considered in \cite{OrlitskyRoche01}. Alice has access to a discrete
random variable $U\in\mathcal{U}$ and Bob has access to $V\in\mathcal{V}$
with a joint distribution $p(u,v)$
and Bob's goal is to
calculate $Z=g(U,V)$ (without distortion). Since
Orlitsky and Roche's approach
is the basis for our solution, we will need to discuss it in some
detail. The key concept is that of independence of points in $\mathcal{U}$. We refer the reader to \cite{OrlitskyRoche01,ElGamalKimBook} for
the definition through graphs; we will provide a more
direct approach. Suppose that $u,u'$ satisfy $\forall v\in\mathcal{V}:g(u,v)=g(u',v)$. In that case, Bob clearly does not need to know if $u$ or $u'$ happened. We
can therefore partition $\mathcal{U}$ into subsets where
$g(\cdot,v)$ is constant, and then Alice just needs to
transmit to Bob in which subset her outcome is. However,
this is too strict a requirement. If $(u,v)$ is impossible,
that is $p(u,v)=0$, we do not need to require $g(u,v)=g(u',v)$ to put $u,u'$ in the same set. This leads to the following
definition.

\begin{definition}
	$u,u'\in\mathcal{U}$ are independent if
	\begin{align*}
  		\forall v\in\mathcal{V}: p(u,v),p(u',v)>0\Rightarrow g(u,v)=g(u',v)
	\end{align*}
\end{definition}
Let $\mathcal{G}$ denote the set of independent sets (i.e., sets where all elements are independent). These no
longer necessarily form a partition,
 and we therefore
let $W$ be a $\mathcal{G}$-valued random variable. We get the
distribution of $W$ by choosing a conditional distribution
$p(w|u)$ where $w$ ranges over all the sets in $\mathcal{G}$ that
contain $u$. Since the outcome $w$ is a subset of $\mathcal{U}$
we can use the notation $u\in w$.
We can restrict $\mathcal{G}$ to the maximal independent sets.
To clarify this concept, a few examples from \cite{OrlitskyRoche01}
are illustrative.

\begin{example}
	If $p(u,v)>0$ for all $u,v$, the maximal independent sets
	are the subsets of $\mathcal{U}$ where $g$ is constant,
	i.e., $\forall v: g(u,v)=g(u',v)$ when $u,u'$ is in the same
	independent set. The maximal independent sets are a partition
	of $\mathcal{U}$ as above and $w$ is a deterministic function of
	$u$. Alice can simply transmit $w$ instead of $u$, and
	with classical Slepian-Wolf coding \cite{CoverBook} the
	rate is $H(W|V)$ (the results in \cite{OrlitskyRoche01}
	shows that this is optimum).
\end{example}

\begin{example}\label{ex:cards}
	Alice and Bob draw a card in $\{1,2,3\}$ from a bag without
replacement. Bob needs to determine who has the largest card.
	In this case $\mathcal{G}=\{\{1,2\},\{2,3\}\}$. It
	is sufficient for Bob to know $w\in\mathcal{G}$. For example,
if $w=\{1,2\}$ and if Bob has $v=1$, he knows that Alice has the
largest card, but if $v\in \{2,3\}$, he knows he has the largest
card.
\end{example}
Orlitsky and Roche \cite{OrlitskyRoche01} show that for any $p(w|u)$ the
rate  
\begin{align}
	R&\geq I(W;U|V)=H(W|V)-H(W|U) \label{eq:Rclassical}
\end{align}
is achievable and further that
\[
  H_G(U|V)\stackrel{\text{def}}{=}\min_{p(w|u)} I(W;U|V)
\]
is optimum. The idea of the achievable rate is as follows.
Alice generates $s$ iid sequences $w^n$ according $p(w)$. Given
a sequence $u^n$ she finds a $w^n$ among the $s$ sequences that
is jointly typical with $u^n$; she then randomly
bins the index into $t$ bins and transmits the bin index
to Bob. If $s$ and $t$ are sufficiently large ($s=2^{n(I(W;U)+\delta)}$), the error
probability can be made to approach zero as $n$ becomes large.

Orlitsky and Roche's proof technique does not directly mix well with
faithful
simulation. 
We propose two ways to overcome this.

In the first approach, instead of measuring $u$ and then finding $w$ as in
the classical case, Alice bases her approach on
measuring $w$ directly. 
That is, Alice uses the POVM
\begin{align}
  \Lambda_{w_A}^A=\sum_{u\in w_A}p_A(w_A|u)\Lambda_u^A \label{eq:lambdaw}
\end{align}
This approach has an analogy in the classical case.
For each outcome $u$, Alice can calculate a \emph{random}
function $w_A(u)$ according to the distribution $p_A(w|u)$. 
It is then clear that the resulting achievable rate with binning (Slepian-Wolf coding) is
$R>H(W_A|V)$, which is worse than (\ref{eq:Rclassical}) except if $W$ is a deterministic function
of $U$, but still better than transmitting $U$ directly,
which gives $R>H(U|V)$.

In this scheme, Bob can also use his possession of the
$B$ system and its entanglement with the $A$ system to help decode Alice's measurements of
$W_A$, as in measurement compression with quantum
side-information \cite{WildeAl2012}. However, Bob
also has to make measurements to compute $g$. In order
to assist in decoding as much as possible, Bob should measure
as gently as possible, just enough to calculate $g$.
So, we define

\begin{definition}
	Let $\mathcal{G}^A$ denote a set of  independent sets 
spanning\footnote{meaning that their union is $\mathcal{U}$} $\mathcal{U}$
for Alice. We define $v,v'\in\mathcal{V}$ to be independent if
\begin{align}
  \forall w\in\mathcal{G}^A:\forall u,u'\in w, p(u,v),p(u',v')>0: g(u,v)=g(u',v') \label{eq:indepB}
\end{align}
\end{definition}
Let $\mathcal{G}^B$ denote a set of 
independent sets spanning $\mathcal{V}$ (not necessarily maximal). Similarly, as for Alice, we can define a 
distribution $p_B(w_B|v)$ and a measurement $\Lambda_{w_B}^B$.

We can now define a function $\tilde g: \mathcal{G}^A\times \mathcal{G}^B\to \mathcal{Z}$ by
$\tilde g(w_A,w_A)=g(u,v)$ for any $u\in w_A,v\in w_B$ with $p(u,v)>0$; the
condition (\ref{eq:indepB}) ensures this is a consistent definition.

This approach leads to the following achievable
capacity region.
\begin{thm}\label{thm:wmeasure}
Let $\mathcal{G}^A,\mathcal{G}^B$ be spanning independent
sets of $\mathcal{U},\mathcal{V}$.
	Let $W_A$ be distributed as $p_A(w|u)$ and
	$W_B$  as $p_B(w|v)$. There exists a faithful (feedback) simulation of $\Lambda_z$
with communication rate $R$ and common randomness rate $S$
if
\begin{align}
  R &\geq I(W_A;RB)-I(W_A;B|W_B)-I(W_A;W_B) \label{eq:WRbound} \\
  R+S &\geq H(W_A|W_B)-I(W_A;B|W_B) \label{eq:WRSbound}
\end{align}
Here $I(W_A;RB)$ is evaluated on the state $\sum_{w_A}|w_A\rangle\langle w_A|\otimes \Tr_A\left\{(I^R\otimes \Lambda_{w_A}^A\otimes I^B)\phi^{RAB}\right\}$
and $I(W_A;B|W_B)$ on 
$\sum_{w_A,w_B}|w_A\rangle\langle w_A|\otimes |w_B\rangle\langle w_B|\otimes\Tr_A\left\{(I^R\otimes \Lambda_{w_A}^A\otimes \Lambda_{w_B}^B)\phi^{RAB}\right\}$.
\end{thm}

The second approach is more similar to \cite{OrlitskyRoche01}. Bob measures (typical) $u^n$
and transmits $w^n$ jointly typical with $w^n$. However,
to enable faithful simulation, we use the following
measurement for $w^n\in T_\delta^{W^n}$
\begin{align*}
  \Lambda^A_{w^n}=\sum_{u^n\in T_\delta^{U^n|w^n} }
  \tilde p(w^n|u^n)\Lambda^A_{u^n}
\end{align*}
where $\tilde p(w^n|u^n)$ is a probability distribution normalized
over typical sequences. This approach
is not really amenable to using $B$ as side information
for decoding.
It results in the following region.

\begin{thm}\label{thm:Ubound}
Let $W^*$ be distributed according to $\arg\min_{p(w|u)} I(W;U|V)$.
There exists a faithful (feedback) simulation of $\Lambda_z$
with communication rate $R$ and common randomness rate $S$
if
\begin{align}
  R&\geq I(U;RB)- I(W^*;V) \label{eq:URbound} \\
  R+S&\geq I(W^*;U|V)=H_G(U|V) \label{eq:URSbound}
\end{align}
Here $I(U;RB)$ is evaluated on the state $\sum_{u}|u\rangle\langle u|\otimes \Tr_A\left\{(I^R\otimes \Lambda_{u}^A\otimes I^B)\phi^{RAB}\right\}$
\end{thm}

\begin{figure}[hbt]
  \includegraphics[width=3.5in]{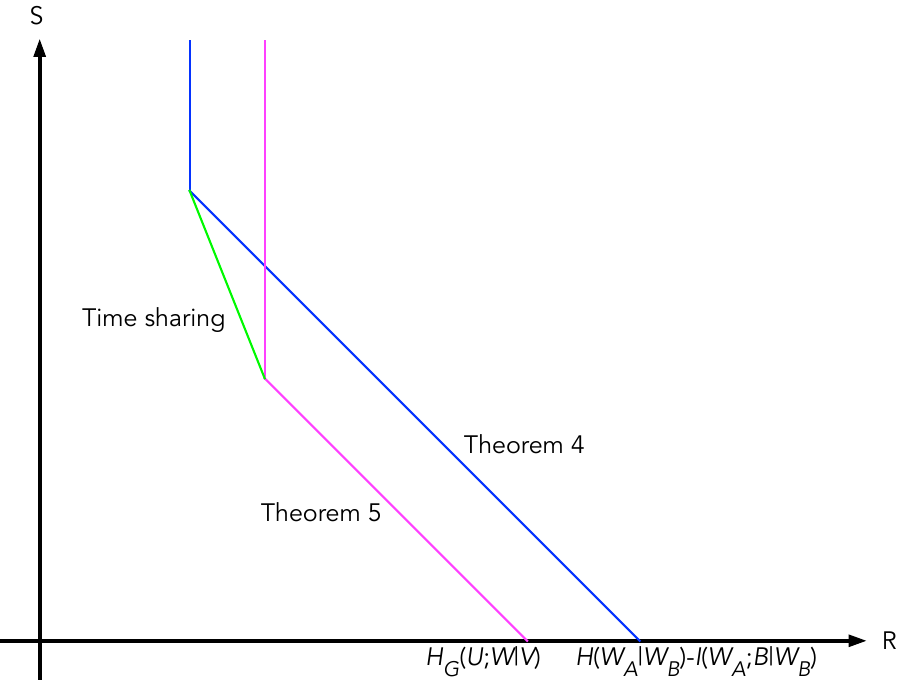}
  \caption{\label{fig:regions} Rate regions for the case $H_G(U;W|V)<H(W_A|W_B)-I(W_A;B|W_B)$.}
\end{figure}

Among these rate regions, (\ref{eq:WRbound}) gives
a lower rate $R$ than (\ref{eq:AtifR}) or (\ref{eq:URbound}): we can simply put $W_A=U$
to equalize the bounds. In general, the optimum
$W_A$ is not a deterministic function of $U$, and
in that case (\ref{eq:WRbound}) is strictly smaller.
On the other hand, either (\ref{eq:WRSbound}) or
(\ref{eq:URSbound}) could be smaller. If, for example,
$W$ is not a deterministic function of $U$ and $A$ and $B$ are separable, (\ref{eq:URSbound}) is
smaller. In the other hand, if $W$ is a deterministic
function of $U$ and $I(W_A;B|W_B)>0$, (\ref{eq:WRSbound})
is smaller. In the latter case, the region of
Theorem \ref{thm:Ubound} is included in that of
Theorem \ref{thm:wmeasure}. Otherwise, both regions
are relevant and can be combined with time-sharing; see Fig. \ref{fig:regions}.

\begin{example}
	We consider a "quantified" version of 
Example \ref{ex:cards}. The ensemble is
$\{\frac 1 6,|u\rangle_A\langle v|_B\}_{u,v\in\{1,2,3\},u\neq v}$
and the measurements $\Lambda^A_u=|u\rangle\langle u|_A$. 
The bound (\ref{eq:WRbound}) gives $R\geq I(W_A;U)-I(W_A;V)$, which
is the same as classical compression (\ref{eq:Rclassical}), and minimized as $0.541$ \cite{OrlitskyRoche01}, whereas (\ref{eq:AtifR}) gives
$R\geq H(U|V)=1$ (the Slepian-Wolf rate). So, (\ref{eq:WRbound})
gives the optimum rate and is strictly better than (\ref{eq:AtifR}).
The bound (\ref{eq:WRSbound}) gives $R+S \geq H(W_A|V)=0.874$, which
is less than (\ref{eq:AtifRS}). However, the rate $0.541$
is achievable without any common randomness simply by using
classical compression (which also simulates $\Lambda_z$), so
(\ref{eq:WRSbound}) cannot be optimum. On the other
hand (\ref{eq:URSbound}) achieves $0.541$.
\end{example}


\section{Proof of Theorem \ref{thm:Ubound}}\label{sec:method1}
Since our result is asymptotic for iid sources, the proof follows
the methodology in \cite{WildeAl2012}. There are also one-shot approaches
to faithful simulation, for example
\cite{Anshu09,Chakraborty22}.

 We will let
$m$ uniform on $[M]$ denote the common randomness.
We first define
\begin{align}
  \hat \rho^A_u &=\frac 1{\Tr\{\Lambda_u^A\rho^A\}}\sqrt{\rho^A}\Lambda_u^A \sqrt{\rho^A} \nonumber\\
  \xi'_{u^n}&=\Pi^\delta_{A^n}\Pi^\delta_{\hat\rho^A|u^n}\hat \rho^A_{u^n} \Pi^\delta_{\hat\rho^A|u^n}\Pi^\delta_{A^n} \label{eq:xiprime}
\end{align}
where $\Pi^\delta_{A^n}$ is the typical
projector for $\rho^A$ and $\Pi^\delta_{\hat\rho^A|u^n}$
is the conditional typical projector for the
ensemble $\{p_U(u),\hat\rho^A_u\}$.

Define the pruned distributions over the typical 
sets
\begin{align}
  \tilde p(w^n)&=\begin{cases}
  	\frac 1 S p(w^n) & w^n\in T_\delta^{W^n} \\
  	0& \text{otherwise}
  \end{cases} \nonumber\\
    \tilde p(u^n|w^n)&=\begin{cases}
  	\frac 1 {S(w^n)} p(u^n|w^n) & u^n\in T_\delta^{U^n|w^n} \\
  	0& \text{otherwise}
  \end{cases} \label{eq:pupruned}
\end{align}
with $S=\sum_{w^n\in T_\delta^{W^n}}p(w^n)$,
 $S(w^n)=\sum_{u^n\in T_\delta^{U^n|w^n}}p_{u^n|w^n}(w^n)$.
Let $\tilde \xi'$ denote the
average of the $\xi_{u^n}$ according to this distribution,
\begin{align*}
  \tilde \xi'&=\sum_{w^n\in T_\delta^{W^n}} \tilde p(w^n)
  \sum_{u^n\in T_\delta^{U^n|w^n}} \tilde p(u^n|w^n)\xi'_{u^n}
\end{align*}
and let $\Pi$ denote the projector onto the subspace
spanned by the eigenvectors of $\tilde\xi'$ with
eigenvalues larger than 
$\epsilon 2^{-n(H(\rho_A)+\delta)}=\epsilon 2^{-n(H(RB)+\delta)}$ and
define
\begin{align*}
  \Omega &= \Pi \tilde\xi' \Pi \nonumber\\
  \xi_{u^n} &= \Pi  \xi'_{u^n} \Pi
\end{align*}
For use later in the proof (in (\ref{eq:TrOmega2})), we will prove some technical properties
of $\Omega$:
\begin{align*}
  \text{rank}(\Omega)&\leq \Tr\Pi \leq \Tr \Pi^\delta_{A^n} \leq 2^{n(H(RB)+\delta)}
\end{align*}
The second inequality is due to the way
$\xi'_{u^n}$ is defined in (\ref{eq:xiprime}) and
the third inequality due to the bound on the dimension
of the typical subspace \cite{WildeBook}.
The implication is that the eigenvalues less
than $\epsilon 2^{-n(H(RB)+\delta)}$ contribute
at most $\epsilon$ to $\Tr\Omega$, and therefore
\begin{align}
  \Tr\Omega&\geq (1-\epsilon)\Tr\tilde\xi'
  \geq (1-\epsilon)^2\Tr \xi'_{u^n}
  \geq (1-\epsilon)^2(1-\epsilon-2\sqrt\epsilon)
  \label{eq:TrOmega}
\end{align}
where we have used that the probability of the
typical set is greater than $1-\epsilon$ and
\cite[(23)]{WildeAl2012}.

For each outcome $m$ of the common randomness we generate
$s$ iid sequences $w^n$ according to $\tilde p(w^n)$, for a total of
$sM$ sequences $w^n(j,m)$ .
We define the POVM used for measurement
simulation as follows
\begin{align*}
  \tilde \Gamma_{j}^{(m)} &= \frac{S S(w^n(j,m))}{(1+\epsilon)s}\nonumber\\
  &\sqrt{\omega^A}^{-1}
  \left(\sum_{u^n\in T_\delta^{U^n|w^n(j,m)}}\tilde p(u^n|w^n(j,m))\xi_{u^n}\right)\sqrt{\omega^A}^{-1}\nonumber\\
\end{align*}

Alice uses binning for the indices $j$ as follows.
For each $m$ let $\Phi^m:[s]\to[t]$ be a random mapping: for each $j=1\dots s$ and $i\in[t]$ is chosen uniformly random. 
She
transmits the bin index of the measurement result to Bob,
who uses this for measurement simulation. In order
for this scheme to work, we need to prove
\begin{itemize}
  \item The set $\Gamma^{(m)}=\{\Gamma^{(m)}_j\}_{j=1}^s$  constitutes a sub-POVM (with high probability).
  \item Upon receiving the bin index $i$, Bob can
      decode the index $j$ and hence $w^n(j,m)$ (with high
      probability).
   \item The resulting measurement faithfully simulates
      $\{\Gamma_z^{AB}\}_z$
\end{itemize}

\subsubsection{The set $\Gamma^{(m)}=\{\Gamma^{(m)}_j\}_{j=1}^s$  constitutes a sub-POVM}

We will show that the set $\Gamma^{(m)}=\{\Gamma^{(m)}_i\}_{i=1}^t$ is a sub-POVM with
high probability. If it is not a sub-POVM we put $\Gamma^{(m)}=\{I\}$. We calculate
\begin{align}
  \sqrt{\omega^A}\sum_{j=1}^s\Gamma_j^{(m)}\sqrt{\omega^A} &= \frac{S }{(1+\epsilon)s}
  \left(\sum_{j=1}^sS(w^n(j,m))\right.\vphantom{\sum_{u^n\in T_\delta^{U^n|w^n(j,m)}}}\nonumber\\
  &\left.\left(\sum_{u^n\in T_\delta^{U^n|w^n(j,m)}}\tilde p(u^n|w^n(j,m))\xi_{u^n}\right)\right) \nonumber\\
  &= \frac{1 }{(1+\epsilon)s}
  \sum_{j=1}^s\tilde\xi_j \label{eq:xijsum}
\end{align}
where
\begin{align*}
  \tilde\xi_j &= S S(w^n(j,m))
  \sum_{u^n\in T_\delta^{U^n|w^n(j,m)}}\tilde p(u^n|w^n(j,m))\xi_{u^n}
\end{align*}
Notice that the $\tilde\xi_j$ are iid, and we can
therefore use the operator Chernoff bound
\cite[Lemma 17.3.1]{WildeBook} to bound
(\ref{eq:xijsum}).
First, we need
\begin{align}
  E[\tilde\xi_j]
  &=\sum_{w^n\in T_\delta^{W^n}}S\tilde p(w^n)
  \sum_{u^n\in T_\delta^{U^n|w^n}}S(w^n)\tilde p(u^n|w^n)\xi_{u^n} \nonumber\\
  &= \sum_{w^n\in T_\delta^{W^n}} p(w^n)
  \sum_{u^n\in T_\delta^{U^n|w^n}} p(u^n|w^n)\xi_{u^n} \nonumber\\
  &\leq \sum_{w^n\in \mathcal{W}^n} p(w^n)
  \sum_{u^n\mathcal{U}^n} p(u^n|w^n)\xi_{u^n} \nonumber\\
  &= \sum_{u^n\mathcal{U}^n} p(u^n)\xi_{u^n}\sum_{w^n\in \mathcal{W}^n} p(w^n|u^n)
   \nonumber\\
  &= \sum_{u^n\mathcal{U}^n} p(u^n)\Pi\Pi^\delta_{A^n}\Pi^\delta_{\hat\rho^A|u^n}\hat \rho^A_{u^n} \Pi^\delta_{\hat\rho^A|u^n}\Pi^\delta_{A^n}\Pi \nonumber\\
  &\leq \sum_{u^n\mathcal{U}^n} p(u^n)\Pi\Pi^\delta_{A^n}\hat \rho^A_{u^n} \Pi^\delta_{A^n}\Pi \nonumber\\
  &= \Pi\Pi^\delta_{A^n} (\rho^A)^{\otimes n} \Pi^\delta_{A^n}\Pi \nonumber\\
  &\leq (\rho^A)^{\otimes n} = \omega^A \label{eq:ExiU}
\end{align}
where the second equality follows from $\Pi^\delta_{\hat\rho^A|u^n}\hat \rho^A_{u^n} \Pi^\delta_{\hat\rho^A|u^n}\leq \hat \rho^A_{u^n}$
and the last equality from $\sum_{u^n\mathcal{U}^n} p(u^n)\hat \rho^A_{u^n} =I$ by its definition.
Thus, $E[\sum_{j=1}^s\Gamma_j^{(m)}]\leq (1+\epsilon)^{-1}I$.
Let $E_m$ be the event that $\sum_{j=1}^s\Gamma_j^{(m)}\leq I$, or, equivalently,
\begin{align*}
  \frac 1 s\sum_{j=1}^s\beta\tilde\xi_j
  &\leq (1+\epsilon)\beta E[\tilde\xi_j]
\end{align*}
where $\beta$ is a scaling factor. In order to use
the operator Chernoff bound, we need $\beta\tilde\xi_j\leq I$:
\begin{align*}
  \beta\tilde\xi_j &= S S(w^n(j,m))
  \sum_{u^n\in T_\delta^{U^n|w^n(j,m)}}\tilde p(u^n|w^n(j,m))\beta\xi_{u^n}\nonumber\\
  &= S S(w^n(j,m))\nonumber\\
  &\sum_{u^n\in T_\delta^{U^n|w^n(j,m)}}\!\!\!\!\!\!\!\!\!\!\!\!\!\!\beta \tilde p(u^n|w^n(j,m))\Pi\Pi^\delta_{A^n}\Pi^\delta_{\hat\rho^A|u^n}\hat \rho^A_{u^n} \Pi^\delta_{\hat\rho^A|u^n}\Pi^\delta_{A^n}\Pi \nonumber\\
  &\leq S S(w^n(j,m))\nonumber\\&
  \sum_{u^n\in T_\delta^{U^n|w^n(j,m)}}\beta \tilde p(u^n|w^n(j,m)) 2^{-n(H(RB|U)-\delta)}\nonumber\\&\times\Pi\Pi^\delta_{A^n}\Pi^\delta_{\hat\rho^A|u^n} \Pi^\delta_{\hat\rho^A|u^n}\Pi^\delta_{A^n}\Pi \nonumber\\
  &\leq S S(w^n(j,m))\beta  2^{-n(H(RB|U)-\delta)} I\nonumber\\&
  \sum_{u^n\in T_\delta^{U^n|w^n(j,m)}}\tilde p(u^n|w^n(j,m))  \nonumber\\
  &\leq S S(w^n(j,m))\beta 2^{-n(H(RB|U)-\delta)} I
\end{align*}
where we have used the equipartition property of
conditional typicality.
Thus, we can choose $\beta = 2^{n(H(RB|U)-\delta)}$.
We also observe that
\begin{align*}
  E[\beta\tilde\xi_j]&=\beta\Omega
  \geq \beta\epsilon 2^{-n(H(RB)+\delta)}
\end{align*}
%
%
The operator Chernoff bound now gives
\begin{align*}
P(E_m^c)&=
P\left(\frac 1 s\sum_{j=1}^s\beta\tilde \xi_{j}> \beta (1+\epsilon)E[\tilde\xi_j]\right) \nonumber\\
&\leq 2\text{rank}(\Pi)
  \exp\left(-\frac{s\epsilon^2(\beta\epsilon 2^{-n(H(RB)+\delta)})}{4\ln 2}\right) \nonumber\\
  &\leq 2\exp\left(-\frac{s\epsilon^3 2^{n(H(RB|U)-\delta)} 2^{-n(H(RB)+\delta)}}{4\ln 2}\right.\nonumber\\ &\left.
  \vphantom{\frac{s\epsilon^3 2^{n(H(RB|U)-\delta)} 2^{-n(H(RB)+\delta)}}{4\ln 2}}+n(H(RB)+\delta)\right)
\end{align*}
If we choose 
\begin{align}
s\geq 2^{n(I(U;RB)+3\delta)}	 \label{eq:scond}
\end{align}
this probability converges to zero. 

The total probability of error then is
\begin{align*}
  P\left(\bigcup_m E_m^c\right)&\leq \sum_mP(E_m^c)\nonumber\\
  &\leq 2M
  \exp\left(-\frac{\epsilon^3 2^{n\delta} }{4\ln 2}+n(H(RB)+\delta)\ln 2\right)
\end{align*}
So, as long as $M\leq O(\exp(n))$, the total error probability
converges to zero.

\subsubsection{Upon receiving the bin index $i$, Bob can
      decode the index $j$ and hence $w_j^n$}
This  only depends on the average number $\frac s t$ of $w_j^n$ in
each bin, not the number of bins. From [Orlitsky] we
know that $\frac s t= 2^{n(I(W;V)+\delta/2)}$ allows
for decoding.

The conclusion is that we need
\begin{align*}
  R&\geq I(U;RB)-I(W;V)
\end{align*}
which is (\ref{eq:URbound}).

\subsubsection{The resulting measurement faithfully simulates
      $\{\Gamma_z^{AB}\}_z$ }
We will evaluate how well the simulation works
under the assumption that $j$ is decoded correctly
and $\Gamma^{(m)}$ is a sub-POVM for all $m$. 
We can define a function $\tilde g^n$ as follows
\begin{align*}
  \tilde g^n(w^n(j,m),v^n) &= g^n(u^n,v^n)\qquad u^n\in 
  T_\delta^{U^n|w^n(j,m)}
\end{align*}
By \cite[Lemma 4]{OrlitskyRoche01}\footnote{which is still
valid with the typicality definition (\ref{eq:TypDef})} this
is well-defined in the sense that it does not
depend on which $u^n\in T_\delta^{U^n|w^n(j,m)}$ is used.

\begin{figure*}[!b]
\normalsize
\hrulefill
\begin{align*}
  d &= \sum_{z^n\in\mathcal{Z}^n}\left\|\sqrt{\omega}
  (\tilde\Lambda_{z^n}^{AB}-\Lambda_{z^n}^{AB})\sqrt{\omega}
  \right\|_1 \nonumber\\
  &\leq \sum_{z^n\notin \mathcal{S}_z}\left\|\sqrt{\omega}
  \Lambda_{z^n}^{AB}\sqrt{\omega}
  \right\|_1+\sum_{z^n\in \mathcal{S}_z}\left\|\sqrt{\omega}
  (\tilde\Lambda_{z^n}^{AB}-\Lambda_{z^n}^{AB})\sqrt{\omega}
  \right\|_1\nonumber\\
  &\leq \epsilon+\sum_{z^n\in \mathcal{S}_z}\left\|
  \sqrt{\omega}\sum_{v^n\in\mathcal{V}^n}\left(\frac 1{1+\epsilon}\sum_{u^n:g^n(u^n,v^n)=z^n} 
  \left(\sqrt{\omega^A}^{-1}\hat p(u^n)\xi_{u^n}\sqrt{\omega^A}^{-1}\right)-\sum_{u^n:g(u^n,v^n)=z^n}\Lambda^A_{u^n}\right)\otimes \Lambda_{v^n}^B\sqrt{\omega}\right\|_1 \nonumber\\
   &\leq \epsilon+\sum_{z^n\in \mathcal{S}_z}\sum_{v^n\in\mathcal{V}^n}\sum_{u^n:g^n(u^n,v^n)=z^n}\left\|
  \sqrt{\omega}\left(\frac 1{1+\epsilon} 
  \left(\sqrt{\omega^A}^{-1}\hat p(u^n)\xi_{u^n}\sqrt{\omega^A}^{-1}\right)-\sum_{u^n:g(u^n,v^n)=z^n}\Lambda^A_{u^n}\right)\otimes \Lambda_{v^n}^B\sqrt{\omega}\right\|_1 \nonumber\\
   &\leq \epsilon+\sum_{v^n\in\mathcal{V}^n}\sum_{u^n\in\mathcal{U}^n}\left\|
  \sqrt{\omega}\left(\frac 1{1+\epsilon} 
  \left(\sqrt{\omega^A}^{-1}\hat p(u^n)\xi_{u^n}\sqrt{\omega^A}^{-1}\right)-\Lambda^A_{u^n}\right)\otimes \Lambda_{v^n}^B\sqrt{\omega}\right\|_1 \nonumber\\
   &\leq \epsilon+\sum_{u^n\in\mathcal{U}^n}\left\|
  \sqrt{\omega^A}\left(\frac 1{1+\epsilon} 
  \left(\sqrt{\omega^A}^{-1}\hat p(u^n)\xi_{u^n}\sqrt{\omega^A}^{-1}\right)-\Lambda^A_{u^n}\right)\sqrt{\omega^A}\right\|_1 \nonumber\\
   &= \epsilon+\sum_{u^n\in\mathcal{U}^n}\left\|
  \frac 1{1+\epsilon} 
  \hat p(u^n)\xi_{u^n}-p(u^n)\hat\rho^A_{u^n}\right\|_1 \nonumber\\
\end{align*}

\end{figure*}

Let $\mathcal{S}_z=\{z^n:\exists j,m,v^n:\tilde g^n(w^n(j,m),v^n)=z^n\}$.
For $z^n\in\mathcal{S}_z$ we consider the following collection of operators
\begin{align}
  \tilde\Lambda^{AB}_{z^n}
  &=\frac 1{M}\sum_m\sum_{v^n\in\mathcal{V}^n} \sum_{j:\tilde g (w^n(j,m),v^n)=z^n}
  \tilde\Gamma^{(m)}_j\otimes \Lambda_{v^n}^B  \nonumber\\
  &= \sum_{v^n\in\mathcal{V}^n} \sum_{m,j:\tilde g (w^n(j,m),v^n)=z^n} 
  \left(\vphantom{\sum_{u^n\in T_\delta^{U^n|w^n(j,m)}}}\frac{S S(w^n(j,m)) }{(1+\epsilon)sMN}\right.\nonumber\\&\left.\sum_{u^n\in T_\delta^{U^n|w^n(j,m)}}\!\!\!\!\sqrt{\omega^A}^{-1}\tilde p(u^n|w^n(j,m))\xi_{u^n}\sqrt{\omega^A}^{-1}\right)\otimes \Lambda_{v^n}^B \nonumber\\
  &= \frac 1{1+\epsilon}\sum_{v^n\in\mathcal{V}^n}\nonumber\\&\sum_{u^n:g^n(u^n,v^n)=z^n} 
  \left(\sqrt{\omega^A}^{-1}\hat p(u^n)\xi_{u^n}\sqrt{\omega^A}^{-1}\right)\otimes \Lambda_{v^n}^B \label{eq:tGzn}
\end{align}
where
\begin{align*}
  \hat p(u^n)=\sum_{w^n\in T_\delta^{W^n}}S(w^n)\tilde p(u^n|w^n)\frac{Sc(w^n)}{sM}
\end{align*}
\begin{align*}
  c(w^n)=|\{(j,m):w^n(j,m)=w^n\}|
\end{align*}
Notice that if $\tilde p(u^n|w^n)>0$, $g^n(u^n,v^n)=\tilde g^n(w^n,v^n)$, which allows us to change
summation to $u^n$ in the last step in (\ref{eq:tGzn}).
If $sM\geq 2^{nI(W;U)+\delta)}$, $P(\mathcal{S}_z)\geq 1-\epsilon$.
Namely, for every $z^n$ that has non-zero probability
there exists $u^n,v^n: z^n=g^n(u^n,v^n)$. And for every typical
$u^n$ there exists $w^n(j,m)$ jointly typical with $u^n$
if $sM\geq 2^{nI(W;U)+\delta)}$ \cite{OrlitskyRoche01}.

We need to evaluate $d = \sum_{z^n\in\mathcal{Z}^n}\left\|\sqrt{\omega}
  (\tilde\Lambda_{z^n}^{AB}-\Lambda_{z^n}^{AB})\sqrt{\omega}
  \right\|_1$, bottom of the page. For the last inequality we used
\cite[Lemma 3]{AtifHeidariSandeep22}.

Let $\mathcal{S}_u=\{u^n\in T_\delta^{U^n}:\hat p(u^n)>0\}$
Then, again using the triangle inequality
\begin{align}
  d&\leq \epsilon+ \sum_{ 
  u^n\notin \mathcal{S}_u}\left\|p(u^n)\hat\rho_{u^n}^A\right\|_1 \nonumber\\&
  +\sum_{ u^n\in \mathcal{S}_u}\left\|\frac 1{1+\epsilon}\hat p(u^n)\xi_{u^n}-p(u^n)\hat\rho_{u^n}^A\right\|_1 \nonumber\\
  &\leq 2\epsilon + \sum_{u^n\in \mathcal{S}_u}\left\|\frac 1{1+\epsilon}\hat p(u^n)\xi_{u^n}-p(u^n)\xi_{u^n} \right.\nonumber\\ &\left.\vphantom{\frac 1{1+\epsilon}}+p(u^n)\xi_{u^n}-p(u^n)\hat\rho_{u^n}^A\right\|_1 \nonumber\\
  &\leq 2\epsilon + \sum_{u^n\in \mathcal{S}_u}\left\|\frac 1{1+\epsilon}\hat p(u^n)\xi_{u^n}-p(u^n)\xi_{u^n}\right\|_1\nonumber\\
  &+ \sum_{u^n\in \mathcal{S}_u}\left\|p(u^n)\xi_{u^n}-p(u^n)\hat\rho_{u^n}^A\right\|_1\nonumber\\ 
  &\leq 2\epsilon + \sum_{u^n\in \mathcal{S}_u}\left|\frac 1{1+\epsilon}\hat p(u^n)-p(u^n)\right|\nonumber\\
  &+ \sum_{u^n\in \mathcal{S}_u}\left\|p(u^n)\xi_{u^n}-p(u^n)\hat\rho_{u^n}^A\right\|_1 \label{eq:d}
\end{align}
We will first bound the last sum here,
\begin{align*}
  d_3 &= \sum_{u^n\in \mathcal{S}_u}\left\|p(u^n)\xi_{u^n}-p(u^n)\hat\rho_{u^n}^A\right\|_1 \nonumber\\
  &\leq \sum_{u^n\in \mathcal{S}_u}p(u^n)\left\|\xi_{u^n}-\xi_{u^n}'\right\|_1 \nonumber\\
  &+\sum_{u^n\in \mathcal{S}_u}p(u^n)\left\|\xi_{u^n}'-\hat\rho_{u^n}^A\right\|_1
\end{align*}
Here
\begin{align*}
\MoveEqLeft
  \sum_{u^n\in \mathcal{S}_u}p(u^n)\left\|\xi_{u^n}-\xi_{u^n}'\right\|_1\nonumber\\ &=\sum_{u^n\in \mathcal{S}_u}p(u^n)\left\|\Pi\xi_{u^n}'\Pi-\xi_{u^n}'\right\|_1 \nonumber\\
&\leq\sum_{u^n\in T_\delta^{U^n}}\tilde p(u^n)\left\|\Pi\xi_{u^n}'\Pi-\xi_{u^n}'\right\|_1 \nonumber\\  &\leq \sum_{w^n\in T_\delta^{W^n}} \tilde p(w^n)
  \sum_{u^n\in T_\delta^{U^n|w^n}} \tilde p(u^n|w^n)\left\|\Pi\xi_{u^n}'\Pi-\xi_{u^n}'\right\|_1 \nonumber\\
  &\leq 2\sqrt{\epsilon'}
\end{align*}
by Gentle Measurement for Ensembles \cite[Lemma 9.4.3]{WildeBook}, as
\begin{align}
  \MoveEqLeft\sum_{w^n\in T_\delta^{W^n}} \tilde p(w^n)
  \sum_{u^n\in T_\delta^{U^n|w^n}}\tilde  p(u^n|w^n)
  \Tr\{\Pi\xi'_{u^n}\Pi\} \nonumber\\
  &=\Tr\Omega\geq 1-\epsilon' \label{eq:TrOmega2}
\end{align}
with $\epsilon'=(1-\epsilon)^2(1-\epsilon-2\sqrt\epsilon)$
by (\ref{eq:TrOmega}). Further, by Gentle Measurement
\cite[Lemma 9.4.2]{WildeBook}
\begin{align*}
  \left\|\xi_{u^n}'-\hat\rho_{u^n}^A\right\|_1
  &=\left\|\Pi^\delta_{A^n}\Pi^\delta_{\hat\rho^A|u^n}\hat \rho^A_{u^n} \Pi^\delta_{\hat\rho^A|u^n}\Pi^\delta_{A^n}-\hat\rho_{u^n}^A\right\|_1\nonumber\\
  &\leq 2\sqrt{\epsilon''}
\end{align*}
as
\begin{align*}
  \MoveEqLeft\Tr\{\Pi^\delta_{A^n}\Pi^\delta_{\hat\rho^A|u^n}\hat \rho^A_{u^n} \Pi^\delta_{\hat\rho^A|u^n}\Pi^\delta_{A^n}\}\nonumber\\&=
  \Tr\{\Pi^\delta_{A^n}\Pi^\delta_{\hat\rho^A|u^n}\hat \rho^A_{u^n} \Pi^\delta_{\hat\rho^A|u^n}\} \nonumber\\
  &\geq \Tr\{\Pi^\delta_{A^n}\hat \rho^A_{u^n} \}
  +\frac 1 2\|\Pi^\delta_{\hat\rho^A|u^n}\hat \rho^A_{u^n} \Pi^\delta_{\hat\rho^A|u^n}-\hat \rho^A_{u^n} \|_1 \nonumber\\
  &\geq 1-\epsilon-\sqrt\epsilon
\end{align*}
Where we have used the trace inequality
$\Tr\{\Lambda\rho\}\geq \Tr\{\Lambda\sigma\}
+\frac 1 2\|\rho-\sigma\|_1$  \cite[Corollary 9.1.1]{WildeBook}.
Thus, we conclude that $d_3\leq \epsilon$ for some
$\epsilon\to 0$.

We now turn to the first sum in (\ref{eq:d}). Notice that
\begin{align*}
  E\left[\hat p(u^n)\right]&=
  \sum_{w^n:(w^n,u^n)\in T_\delta^{(W^n,U^n)}}p(u^n|w^n)p(w^n)\nonumber\\&=
  \sum_{w^n:(w^n,u^n)\in T_\delta^{(W^n,U^n)}}p(w^n|u^n)p(u^n)\nonumber\\&=
  \gamma(u^n) p(u^n)
\end{align*}
where
\begin{align*}
  \exists N\forall n>N\forall u^n\in T_\delta^{U^n}:
  1\geq \gamma(u^n)>1-\epsilon
\end{align*}
as each typical $u^n$ has a minimum empirical frequency of each symbol $u\in\mathcal{U}$.
We only consider $u^n$ that are joint typical with some
typical
$w^n$, and for those we can lower bound this probability
as follows
\begin{align*}
  \MoveEqLeft\sum_{w^n: (w^n,u^n)\in T_\delta^{(W^n,U^n)}}p(u^n|w^n)p(w^n) \nonumber\\
  &\geq 2^{-n(H(U|W)+\delta)}\sum_{w^n:(w^n,u^n)\in T_\delta^{(W^n,U^n)}}p(w^n) \nonumber\\
  &\geq 2^{-n(H(U|W)+\delta)}2^{n(H(W|U)-\delta)}2^{-n(H(W)+\delta)} \nonumber\\
  &= 2^{-n(H(U|W)+\delta)}
  2^{-n(I(U;W)+2\delta)}
\end{align*}
We can also rewrite
\begin{align*}
  \hat p(u^n)&\stackrel{\text{def}}{=} \sum_{w^n\in T_\delta^{W^n}}S(w^n)\tilde p(u^n|w^n)\frac{Sc(w^n)}{sM} \nonumber\\
  &= \frac S{sM}
  \sum_{w^n:(w^n,u^n)\in T_\delta^{(W^n,U^n)}}S(w^n)\tilde p(u^n|w^n) \nonumber\\&
  \sum_{m,j}I_{w^n(j,m)=w^n}\nonumber\\
  &= \frac S{sM}
  \sum_{w^n:(w^n,u^n)\in T_\delta^{(W^n,U^n)}}\!\!\!\!\!\!\!\!\!\!\!\!\!\!\!\!
  p(u^n|w(j,m))I_{w^n(j,m)=w^n} \nonumber\\
  &= \frac S{sM}\sum_{m,j}S(w^n)\tilde p(u^n|w^n(j,m))
\end{align*}

We will again use the operator Chernoff bound.
To that end,
let $P$ be the $|T_\delta^{U^n}|\times |T_\delta^{U^n}|$ diagonal matrix of $p(u^n)$ for $u^n\in T_\delta^{U^n}$ ,
and the $C$ be the diagonal matrix of the empirical measures
$\hat p(u^n)$ and $\gamma$ the diagonal matrix
of the $\gamma(u^n)$. By the above,
\begin{align*}
  C & = \frac 1{sM}\sum_{m,j}C_{m,j}\nonumber\\
  E[C]&=\gamma P \nonumber \\
  \gamma P&\geq 2^{-n(H(U|W)+\delta)}
  2^{-n(I(U;W)+\delta)}I
\end{align*}
where $C_{m,j}$ is the diagonal matrix of $SS(w^n)\tilde p(u^n|w^n(j,m))$.
Now
\begin{align*}
  \MoveEqLeft\sum_{u^n\in \mathcal{S}_u}\left|\frac 1{1+\epsilon}\hat p(u^n)-p(u^n)\right|\nonumber\\ &= \left\|\frac 1{1+\epsilon}C-P\right\|_1 = \frac 1{1+\epsilon}\left\|C-(1+\epsilon)P\right\|_1 \nonumber\\
  &\leq \frac 1{1+\epsilon}(\left\|\epsilon P\right\|_1+\left\|C-P\right\|_1) \nonumber\\
  &= \epsilon+\left\|C-\gamma P+(\gamma-1)P\right\|_1 \nonumber\\
  &\leq \epsilon+\left\|C-\gamma P\right\|_1+\left\|(\gamma-1)P\right\|_1 \nonumber\\& \leq \epsilon+\left\|C-\gamma P\right\|_1+\left\|\epsilon P\right\|_1 \nonumber\\
  &\leq 2\epsilon+\left\|C-\gamma P\right\|_1
\end{align*}
We want to show using the operator Chernoff bound that with high probability the event $E_0$ happens:
\begin{align*}
  (1-\epsilon)\gamma P\leq C\leq (1+\epsilon)\gamma P
\end{align*}
which means $\sum_{u^n\in \mathcal{S}_u}\left|\frac 1{1+\epsilon}\hat p(u^n)-p(u^n)\right|\leq 3\epsilon$ with high probability.
By strong conditional typicality $2^{n(H(U|W)-\delta)}p(u^n|w^n(j,m))\leq 1$, or
\begin{align*}
  2^{n(H(U|W)-\delta)}C_{m.j}\leq I
\end{align*}
By the operator Chernoff bound,
\begin{align*}
  \MoveEqLeft P(E_0^c)\leq 2|T_\delta^{U^n}|\nonumber\\&\exp\left(-\frac{sM\epsilon^2 2^{-n(H(U|W)+\delta)}
  2^{-n(I(U;W)+\delta)}2^{n(H(U|W)-\delta)}}{4\ln 2}\right) \nonumber\\
  &\leq 2\cdot 2^{n(H(U)+\delta)}\nonumber\\&\exp\left(-\frac{sM\epsilon^2 2^{-n(H(U|W)+\delta)}
  2^{-n(I(U;W)+\delta)}2^{n(H(U|W)-\delta)}}{4\ln 2}\right) 
\end{align*}
So, if we choose
\begin{align}
  sM&\geq 2^{n(I(U;W)+4\delta)} \label{eq:sMbound}
\end{align}
$P(E_0^c)$ will go to zero.

All of these bounds were under the assumption that no error
occurs. But if an error occurs, the contribution to
$d$ is at most 1, and the total error probability converges
to zero with $n$. 

From (\ref{eq:sMbound}) we get
\begin{align*}
  tM &\geq sM\frac t s \geq 2^{n(I(U;W)+4\delta)}
  2^{-n(I(W;V)+\delta/2)}
\end{align*}
which gives (\ref{eq:URSbound}).

\section{Proof of Theorem \ref{thm:wmeasure}}
We first define
\begin{align}
  \hat \rho^A_w &=\frac 1{\Tr\{\Lambda_w^A\rho^A\}}\sqrt{\rho^A}\Lambda_{w_A}^A \sqrt{\rho^A} \nonumber\\
  \xi'_{w_A^n}&=\Pi^\delta_{A^n}\Pi^\delta_{\hat\rho^A|{w^n}}\hat \rho^A_{w^n} \Pi^\delta_{\hat\rho^A|{w^n}}\Pi^\delta_{A^n} \label{eq:xiwprime}\\
  \xi' &= \sum_{w^n}\tilde p_A(w^n)\xi'_{w_A^n}
  \label{eq:xiw}
\end{align}
where $\Pi^\delta_{A^n}$ is the typical
projector for $\rho^A$ and $\Pi^\delta_{\hat\rho^A|W^n}$
is the conditional typical projector for the
ensemble $\{p_A(w),\hat\rho^A_w\}$
\begin{align}
  \tilde p_{A}(w^n)=\begin{cases}
  	\frac 1 S p_{A}(w^n) & w^n\in T_\delta^{W_A^n} \\
  	0& \text{otherwise}
  \end{cases} \label{eq:pwpruned}
\end{align}
with $S=\sum_{w^n\in T_\delta^{W_A^n}}p_{A}(w^n)$.

The essential difference from Theorem \ref{thm:Ubound} is
that the fundamental measurement is of conditional
typicality with $w^n$ in (\ref{eq:xiwprime}) whereas
in (\ref{eq:xiprime}) it is conditional
typicality with $u^n$.

We let $\Pi$ be the projector onto the eigenvectors of $\xi'$ greater than
$\epsilon 2^{-n(H(\rho^A)+\delta)}=\epsilon 2^{-n(H(RB)+\delta)}$ and
define
\begin{align}
  \Omega &= \Pi \xi' \Pi \nonumber\\
  \xi_{w_A^n} &= \Pi  \xi'_{w_A^n} \Pi \label{eq:xiwan}
\end{align}
As in the proof of Theorem \ref{thm:Ubound} we have
\begin{align}
  \Tr\Omega&\geq (1-\epsilon)(1-\epsilon-2\sqrt\epsilon)
  \label{eq:TrOmegaw}
\end{align}

We generate random $w_A^n(j,m),j\in [s],m\in [M]$  according
to $\tilde p_{A}$.
For $j\in[s]$ we define the operators
\begin{align}
  \Gamma^{(m)}_j &= \frac{S}{(1+\epsilon)s}\sqrt{\omega^A}^{-1}
  \xi_{w_A^n(j,m)}\sqrt{\omega^A}^{-1}
  \label{eq:Gammajw}
\end{align}
The $s$ possible outcomes are randomly binned into $t$ bins, and
the bin index is transmitted to Bob.


In order
for this scheme to work, we need to prove
\begin{enumerate}
  \item The set $\Gamma^{(m)}=\{\Gamma^{(m)}_j\}_{j=1}^s$  constitutes a sub-POVM (with high probability).
  \item Upon receiving the bin index $i$ (and knowing $m$), Bob can
      decode  $w_A^n(j,m)$ (with high
      probability).
   \item The resulting measurement faithfully simulates
      $\{\Gamma_z^{AB}\}_z$
\end{enumerate}

\subsubsection{The set $\Gamma^{(m)}=\{\Gamma^{(m)}_j\}_{j=1}^s$  constitutes a sub-POVM}

We will show that the set $\Gamma^{(m)}=\{\Gamma^{(m)}_j\}_{j=1}^s$ is a sub-POVM with
high probability. If it is not a sub-POVM we put $\Gamma^{(m)}=\{I\}$. 
The proof is very similar to the proof
for Theorem \ref{thm:Ubound} and to \cite{WildeAl2012}, so
we will only outline it.

We calculate
\begin{align*}
  \sqrt{\omega^A}\sum_{j=1}^s\Gamma_j^{(m)}\sqrt{\omega^A} 
  &= \frac{S}{(1+\epsilon)}
  \left(\frac 1 s\sum_{j=1}^s\xi_{w_A^n(j,m)}\right)
\end{align*}
where each $w^n(j,m)$ is chosen independently according to $p_{\tilde W^n}$.
Similarly to (\ref{eq:ExiU}) we have the following
\begin{align}
	SE[\xi_{w_A^n(j,m)}] &\leq \omega^A \label{eq:SOomegaw}
\end{align}
Let $E_m$ be the event that
\begin{align*}
\frac 1 s\sum_{j=1}^s\beta\xi_{w_A^n(j,m)}\leq \beta\Omega (1+\epsilon)
\end{align*}
for some scaling factor $\beta$. By (\ref{eq:SOomegaw})
this event is equivalent to $\sum_{j=1}^s\Gamma_j^{(m)}\leq I$, i.e., that $\Gamma^{(m)}$ is 
a sub-POVM. We will show that $E_m$ happens
with high probability using the operator Chernoff bound
\cite[Lemma 17.3.1]{WildeBook}. We notice that by (\ref{eq:SOomegaw}) and the definition
of $\Pi$, $E[\beta\xi_{w_A^n(j,m)}]=\beta\Omega
\geq \beta\epsilon 2^{-n(H(RB)+\delta)}\Pi$. Furthermore,
\begin{align*}
\beta\xi_{w_A^n}&=\beta\Pi\Pi^n_{\rho^A,\delta}\Pi^n_{\hat\rho^A_{w^n},\delta}\hat \rho^A_{w^n} 
  \Pi^n_{\hat\rho^A_{w^n},\delta}\Pi^n_{\rho^A,\delta}\Pi \nonumber\\
  &\leq \beta\Pi\Pi^n_{\rho^A,\delta}2^{-n(H(RB|W_A)-\delta)}\Pi^n_{\hat\rho^A_{w^n},\delta}\Pi^n_{\rho^A,\delta}\Pi \nonumber\\
  &\leq \Pi 
\end{align*}
when $\beta= 2^{n(H(RB|W_A)-\delta)}$. The first inequality
follows from properties of conditional quantum typicality
\cite{WildeBook}. Then by the operator Chernoff bound 
\begin{align*}
P(E_m^c) &=
P\left(\frac 1 s\sum_{j=1}^s\beta\xi_{w_A^n(j,m)}> \beta\Omega (1+\epsilon)\right)\nonumber\\
&\leq 2\text{rank}(\Pi)
  \exp\left(-\frac{s\epsilon^2\beta\epsilon 2^{-n(H(RB)+\delta)}}{4\ln 2}\right) \nonumber\\
&\leq 2
  \exp\left(-\frac{s\epsilon^32^{n(H(RB|W_A)-\delta)} 2^{-n(H(RB)+\delta)}}{4\ln 2}\right.\nonumber\\&\left.
  \vphantom{\frac{s\epsilon^32^{n(H(RB|W_A)-\delta)} 2^{-n(H(RB)+\delta)}}{4\ln 2}}+n(H(RB)+\delta)\ln 2\right)
\end{align*}
Then
with 
\begin{align}
  s=2^{n(I(W_A;RB)+3\delta)} \label{eq:sw}
\end{align}
the error probability
goes to zero. 

\subsubsection{Bob can decode $w_A^n$}
Bob measures $w_B^n$, by using the POVM
\begin{align}
  \Lambda_{w_B^n}^B=\sum_{v^n\in w_B^n}p_B^n(w_B^n|v^n)\Lambda_{v^n}^A 
\end{align}
for $w_B^n\in T_\delta^{W_B^n}$, supplemented with
$I-\sum_{w_B^n\in T_\delta^{W_B^n}}\Lambda_{w_B^n}^B$. For
the latter outcome, an error is declared, with a probability less than $\epsilon$.
Bob also receives the bin index $i$. He then looks in bin $i$ for $w^n_A$
that are jointly typical with $w_B^n$; call this set
$\mathcal{S}^{(m)}(i,w_B^n)$. Let the index $k$ enumerate $\mathcal{S}^{(m)}(i,w_B^n)$.
The post-measurement states are
\begin{align}
  \tilde\rho^{B^n}_{w_B^n,w_A^n}
  &=\frac 1{p(i,k,m,w_B^n)}\Tr_{A^n}\left\{((\Gamma_{i,k}^{(m)})^A\otimes \Lambda_{w_B^n}^B)(\rho^{AB})^{\otimes n}\right\} \label{eq:ABstate}
\end{align}
with probabilities
\begin{align*}
  p(i,k,m,w_B^n) &= \Tr\left\{((\Gamma_{i,k}^{(m)})^A\otimes \Lambda_{w_B^n}^B)(\rho^{AB})^{\otimes n}\right\}
\end{align*}
If Alice had done the ideal measurement, the post-measurement
state would have been
\begin{align}
  \rho^{B^n}_{w_B^n,w_A^n}
  &=\frac 1{p(w_A^n,w_B^n)}\Tr_{A^n}\left\{(\Lambda^A_{w_A^n}\otimes \Lambda_{w_B^n}^B)(\rho^{AB})^{\otimes n}\right\} \label{eq:idealstate}
\end{align}
with probabilities
\begin{align*}
  p(w_A^n,w_B^n) &= \Tr\left\{(\Lambda^A_{w_A^n}\otimes \Lambda_{w_B^n}^B)(\rho^{AB})^{\otimes n}\right\}
\end{align*}
We  consider the conditional
typicality projectors for the tensor-power state (\ref{eq:idealstate})
\begin{align}
  \Pi_{i,k,m}&=\Pi^\delta_{B^n|w_A^n(i,k,m),w^n_B}
  \label{eq:Pijkm}
\end{align}
applied to the actual state (\ref{eq:ABstate}).
Bob first uses a conditional typical projector $\Pi^{\delta}_{B^n|w_B^n}$ followed by
sequential decoding with $\{\Pi_{i,k,m},\hat \Pi_{i,k,m}\}$, where
$\hat \Pi_{i,k,m}=I-\Pi_{i,k,m}$. The probability of
correct decoding of the $k$-th message is
\begin{align*}
  P_c &= \Tr\{\acute \Pi_{i,k,m}\tilde\rho^{B^n}_{w_B^n,w_A^n(i,k,m)}\grave \Pi_{i,k,m}\} \nonumber\\
  \acute \Pi_{i,k,m} &= \Pi_{i,k,m}\hat\Pi_{i,k-1,m}\cdots\hat\Pi_{i,1,m}\Pi^\delta_{B^n|w_B^n} \nonumber\\
  \grave \Pi_{i,k,m}&=\Pi^\delta_{B^n|w_B^n}\hat\Pi_{i,1,m}\hat\Pi_{i,k-1,m}\cdots\Pi_{i,k,m}
\end{align*}
The error probability is
\begin{align}
  P_e&=1-E\left[\frac 1 M\sum_{w_B^n\in T_\delta^{W_B^n}}\sum_{i,k,m}p(i,k,m,w_B^n)\right.\nonumber\\
  &\left.\vphantom{\frac 1 M\sum_{w_B^n\in T_\delta^{W_B^n}}}\Tr\{\acute \Pi_{i,k,m}\tilde\rho^{B^n}_{w_B^n,w_A^n(i,k,m)}\grave \Pi_{i,k,m}\}\right] 
  \label{eq.Pe1}
\end{align}
\begin{align}
  &=1-E\left[\frac 1 M\sum_{w_B^n\in T_\delta^{W_B^n}}\sum_{i,k,m}p(i,k,m,w_B^n)\right.\nonumber\\
  &\left.\vphantom{\frac 1 M\sum_{w_B^n\in T_\delta^{W_B^n}}}\Tr\{\Upsilon_{i,k,m}\tilde\rho^{B^n}_{w_B^n,w_A^n(i,k,m)}\}\right] \label{eq:Pea}
\end{align}
where the equality is due to the rotation invariance of the trace with
\begin{align*}
  \Upsilon_{i,k,m} &= \Pi^\delta_{B^n|w_B^n}\hat\Pi_{i,1,m}\hat\Pi_{i,k-1,m}\cdots\Pi_{i,k,m} \nonumber\\
  &\times \Pi_{i,k,m}\hat\Pi_{i,k-1,m}\cdots\hat\Pi_{i,1,m}\Pi^\delta_{B^n|w_B^n}
\end{align*}
The outer sum in (\ref{eq.Pe1}) is explicitly for $w_B^n\in T_\delta^{W_B^n}$ and the expectation is both
over the random choice of $w_A^n(j,m)$ and the random
binning.

The first step in the proof is to show that measuring on
the states $\tilde\rho^{B^n}_{w_B^n,w_A^n}$ is almost
equivalent to measuring on the tensor product states
$\rho^{B^n}_{w_B^n,w_A^n}$, which enables using 
typicality methods. 
To that end, we would like to move $\Upsilon_{i,k,m}$
outside the summation over $i,k,m$. We therefore define
\begin{align}
  \Upsilon_{w_A^n,w_B^n}&=\arg\min_{\Upsilon_{i,k,m}:w_A^n(i,k,m)=w_A^n}\Tr\{\Upsilon_{i,k,m}\tilde\rho^{B^n}_{w_B^n,w_A^n(i,k,m)}\}
  \label{eq.minwA}
\end{align}
And $\Upsilon_{w_A^n,w_B^n}=I$ if there is none.
We can then write
\begin{align}
  P_e &= \frac 1 M\sum_{w_A^n\in T_\delta^{W_A^n},w_B^n\in T_\delta^{W_B^n}}\!\!\!\!\!\!
  E\left[\sum_{i,k,m}p(i,k,m,w_B^n)I_{w_A^n=w_A^n(i,k,m)}\right.\nonumber\\
  &\left.\vphantom{\sum_{i,k,m}}\Tr\{(I-\Upsilon_{i,k,m})\tilde\rho^{B^n}_{w_B^n,w_A^n(i,k,m)}\}\right] \nonumber\\
  &\leq \frac 1 M\sum_{w_A^n\in T_\delta^{W_A^n},w_B^n\in T_\delta^{W_B^n}}
  E\left[\Tr\left\{(I-\Upsilon_{w_A^n,w_B^n})\vphantom{\sum_{i,k,m}}\right.\right.\nonumber\\
  &\left.\left.\sum_{i,k,m}p(i,k,m,w_B^n) I_{w_A^n=w_A^n(i,k,m)}
  \tilde\rho^{B^n}_{w_B^n,w_A^n(i,k,m)}\right\}\right]\nonumber\\
  &\leq \!\!\!\!\sum_{w_A^n\in T_\delta^{W_A^n},w_B^n\in T_\delta^{W_B^n}}\!\!\!\!\!\!\!\!\!
  E\left[\Tr\left\{(I-\Upsilon_{w_A^n,w_B^n})p(w_A^n,w_B^n)\rho^{B^n}_{w_A^n,w_B^n}\right\}\right] \nonumber\\
  &+E\left[\sum_{w_A^n\in T_\delta^{W_A^n},w_B^n\in T_\delta^{W_B^n}}\left\|\vphantom{\sum_{i,k,m}}p(w_A^n,w_B^n)\rho^{B^n}_{w_A^n,w_B^n}\right.\right.\nonumber\\
  &\left.\left.-\!\!\!\!\sum_{i,k,m:k\in\mathcal{S}^{(m)}(i,w_B^n)}\!\!\!\!\!\!\!\!\!\!\!\!\!\!\!\!p(i,k,m,w_B^n)I_{w_A^n=w_A^n(i,k,m)}
  \tilde\rho^{B^n}_{w_B^n,w_A^n(i,k,m)}\right\|_1\right]
  \label{eq.Pe2}
\end{align}
with the second inequality due to the trace inequality,
$\Tr\{\Lambda\rho\}\leq \Tr\{\Lambda\sigma\}+\|\rho-\sigma\|_1$. The second term in (\ref{eq.Pe2}) is
equivalent to the faithful simulation criterion, which
will be shown in Section \ref{sec:faithful} to
be less than $\epsilon$. We will bound
the first term of (\ref{eq.Pe2}). We again use
rotational invariance of  trace to rewrite it
in the form (\ref{eq.Pe1}) as
\begin{align*}
    P_e&\leq1-E\left[\sum_{w_B^n\in T_\delta^{W_B^n}}\sum_{i,k,m\in\mathcal{S}'(w_B^n)}p(i,k,m,w_B^n)\right.\nonumber\\
    &\left.\vphantom{\sum_{w^B_n}}\Tr\{\acute\Pi_{i,k,m}\rho^{B^n}_{w_B^n,w_A^n(i,k,m)}\grave\Pi_{i,k,m}\}\right] +\epsilon
\end{align*}
where $\mathcal{S}'(w_B^n)$ are the indices that achieve the minimum 
in (\ref{eq.minwA}). Notice that
\begin{align*}
  1&=\Tr \tilde\rho^{B^n}_{w_B^n,w_A^n(i,k,m)}
  =\Tr\{\Pi^\delta_{B^n|w_B^n}\tilde\rho^{B^n}_{w_B^n,w_A^n(i,k,m)}\}\nonumber\\
  &+\Tr\{\hat\Pi^\delta_{B^n|w_B^n}\rho^{B^n}_{w_B^n,w_A^n(i,k,m)}\} \nonumber\\
  &\leq \Tr\{\Pi^\delta_{B^n|w_B^n}\rho^{B^n}_{w_B^n,w_A^n(i,k,m)}\Pi^\delta_{B^n|w_B^n}\}+\epsilon
\end{align*}
Then by the non-commutative union bound \cite[Section 16.6]{WildeBook}
\begin{align}
  P_e &\leq 2\left(\sum_{w_B^n\in T_\delta^{W_B^n}}E\left[\sum_{i,k,m\in\mathcal{S}'(w_B^n)}p(i,k,m,w_B^n)
  \right.\right.\nonumber\\
  &\left(\vphantom{\sum_{l=1}^{k-1}}
  \Tr\{\hat\Pi_{i,k,m}\Pi^\delta_{B^n|w_B^n}\rho^{B^n}_{w_B^n,w_A^n(i,k,m)}\Pi^\delta_{B^n|w_B^n}\}\right.\nonumber\\
  &\left.\left.\left.\!\!+\sum_{l=1}^{k-1}\Tr\{\Pi_{i,l,m}\Pi^\delta_{B^n|w_B^n}\rho^{B^n}_{w_B^n,w_A^n(i,k,m)}\Pi^\delta_{B^n|w_B^n}\}\right)\right]\right)^{1/2}\!\!\!\!+\epsilon
  \label{eq:ncub}
\end{align}
The first term in the inner parentheses can be bounded by the trace inequality
\begin{align*}
  T_1 &=\Tr\{\hat\Pi_{i,k,m}\Pi^\delta_{B^n|w_B^n}\rho^{B^n}_{w_B^n,w_A^n(i,k,m)}\Pi^\delta_{B^n|w_B^n}\} \nonumber\\
  &\leq \Tr\{\hat\Pi_{i,k,m}\rho^{B^n}_{w_B^n,w_A^n(i,k,m)}\}\nonumber\\
  &+\|\Pi^\delta_{B^n|w_B^n}\rho^{B^n}_{w_B^n,w_A^n(i,k,m)}\Pi^\delta_{B^n|w_B^n}
  -\rho^{B^n}_{w_B^n,w_A^n(i,k,m)}\|_1 \nonumber\\
  &\leq \epsilon+2\sqrt\epsilon
\end{align*}
where the $\epsilon$ follows from conditional typicality, and
the second from Gentle Measurement
\cite[Lemma 9.4.2]{WildeBook} as conditional typicality gives
$\Tr\{\Pi^\delta_{B^n|w_B^n}\rho^{B^n}_{w_B^n,w_A^n(i,k,m)}\Pi^\delta_{B^n|w_B^n}\}\geq 1-\epsilon$.
By replacing the summation $l=1,\ldots,k-1$ with all $l\neq k$ the second term in the inner parentheses in (\ref{eq:ncub}) can be bounded by
\begin{align*}
  T_2 
  &\leq \sum_{w_B^n\in T_\delta^{W_B^n}}E\left[\sum_{i,k,m\in\mathcal{S}'(w_B^n)}p(i,k,m,w_B^n)
  \right.\nonumber\\
  &\left.\sum_{l\in\mathcal{S}^{(m)}(i,w_B^n),l\neq k}
  \Tr\{\Pi_{i,l,m}\Pi^\delta_{B^n|w_B^n}\rho^{B^n}_{w_B^n,w_A^n(i,k,m)}\Pi^\delta_{B^n|w_B^n}\}\right] 
\end{align*}
The first summation is over a restricted set of the indices where $w_A^n,w_B^n$ are jointly typical; the sum
does not decrease if we instead sum over all indices where
$w_A^n,w_B^n$ are jointly typical. 
Summing
over all $i$ (bin number) and $k$ ( index) is equivalent
to summing over all $j\in[s]$ where $w_A^n,w_B^n$
are jointly typical; we denote this set $\mathcal{T}(w_B^n)$. Similarly for the second summation, so that
\begin{align*}
  T_2 &\leq  \sum_{w_B^n\in T_\delta^{W_B^n}}E\left[\sum_{j\in\mathcal{T}(w_B^n),m}p(w_A^n(j,m),w_B^n)\sum_{j'\in\mathcal{T}(w_B^n): j'\neq j}\!\!\!I_{B_j=B_{j'}}\right.\nonumber\\
  &\left.\vphantom{\sum_{w^B_n}} 
  \Tr\{\Pi^\delta_{B^n|w_A^n(j',m),w_B^n}\Pi^\delta_{B^n|w_B^n}\rho^{B^n}_{w_B^n,w_A^n(j,m)}\Pi^\delta_{B^n|w_B^n}\}\right] \nonumber\\
\end{align*}
Here we  move the expectation over  random
binning inside the sum and notice that
$E[I_{B_j=B_{j'}}]=P(B_j=B_{j'})=\frac 1 t$, where $t$ is
the number of bins, so that
\begin{align*}
  T_2  &\leq \frac 1 t\sum_{w_B^n\in T_\delta^{W_B^n}}E\left[\sum_{j\in\mathcal{T}(w_B^n),m}p(w_A^n(j,m),w_B^n)\sum_{j'\in\mathcal{T}(w_B^n):j'\neq j}\right.\nonumber\\
  &\left.\vphantom{\sum_{j'\in\mathcal{T}(w_B^n):j'\neq j}}
  \Tr\{\Pi^\delta_{B^n|w_A^n(j',m),w_B^n}\Pi^\delta_{B^n|w_B^n}\rho^{B^n}_{w_B^n,w_A^n(j,m)}\Pi^\delta_{B^n|w_B^n}\}\right] \nonumber\\
  \end{align*}
  where the expectation now is over the random choice of $w_A^n(j,m)$.
  By classical typicality, $p(w_A^n(j,m),w_B^n)\leq 2^{-n(H(W_A,W_B)-\delta)}$, so with
  $T_2=\frac 1 t T_2'2^{-n(H(W_A,W_B)-\delta)}$
  \begin{align*}
  T_2' &\leq \sum_{w_B^n\in T_\delta^{W_B^n}}E\left[\sum_{j\in\mathcal{T}(w_B^n),m}
  \sum_{j'\in\mathcal{T}(w_B^n):j'\neq j}\right.\nonumber\\
  &\left.\vphantom{\sum_{w^B_n}}\Tr\{\Pi^\delta_{B^n|w_A^n(j',m),w_B^n}\Pi^\delta_{B^n|w_B^n}\rho^{B^n}_{w_B^n,w_A^n(j,m)}\Pi^\delta_{B^n|w_B^n}\}\right] \nonumber\\
  &= \sum_{w_B^n\in T_\delta^{W_B^n}}\sum_{j\in\mathcal{T}(w_B^n),m}
  \sum_{j'\in\mathcal{T}(w_B^n):j'\neq j}\nonumber\\
  &\Tr\{E[\Pi^\delta_{B^n|w_A^n(j',m),w_B^n}]\Pi^\delta_{B^n|w_B^n}E[\rho^{B^n}_{w_B^n,w_A^n(j,m)}]\Pi^\delta_{B^n|w_B^n}\}\nonumber\\
\end{align*}
 where we used that $w_A^n(j,m)$ and $w_A^n(j',m)$ are
  chosen independently. We now use that
  $E[\rho^{B^n}_{w_B^n,w_A^n(j,m)}]\leq\frac 1{1-\epsilon}\rho^{B^n}_{w_B^n}$ \cite{WildeAl2012}
 as the expectation is over typical $w_A^n(j,m)$: \begin{align*}  
  T_2' &\leq \frac 1{1-\epsilon}\sum_{w_B^n\in T_\delta^{W_B^n}}\sum_{j\in\mathcal{T}(w_B^n),m}
  \sum_{j'\in\mathcal{T}(w_B^n):j'\neq j}\nonumber\\
  &\Tr\{E[\Pi_{w_A^n(j',m),w_B^n}]\Pi^\delta_{B^n|w_B^n}\rho^{B^n}_{w_B^n}\Pi^\delta_{B^n|w_B^n}\}
  \end{align*}
Now, by conditional quantum
 typicality
  \begin{align*}
  T_2'&\leq \frac 1 {(1-\epsilon)}2^{-n(H(B|W_B)-\delta)}\sum_{w_B^n\in T_\delta^{W_B^n}}\sum_{j\in\mathcal{T}(w_B^n),m}
  \sum_{j'\in\mathcal{T}(w_B^n):j'\neq j}\nonumber\\
  &\Tr\{E[\Pi_{w_A^n(j',m),w_B^n}]\Pi^\delta_{B^n|w_B^n}\} \nonumber\\
  &\leq \frac 1 {(1-\epsilon)}2^{-n(H(B|W_B)-\delta)}2^{n(H(B|W_A,W_B)+\delta)}\nonumber\\
  &\sum_{w_B^n\in T_\delta^{W_B^n}}\sum_{j\in\mathcal{T}(w_B^n),m}
  \sum_{j'\in\mathcal{T}(w_B^n):j'\neq j}1
\end{align*}
The sum can be bounded using classical
typicality, and putting it together, we get
\begin{align*}
  T_2&\leq \frac 1 {t(1-\epsilon)}2^{-n(H(W_A,W_B)-\delta)}2^{-n(H(B|W_B)-\delta)}\nonumber\\
  &2^{n(H(B|W_A,W_B)+\delta)}2^{n(H(W_B)+\delta)}\nonumber\\
  &Ms2^{-n(I(W_A;W_B)-\delta)}s2^{-n(I(W_A;W_B)-\delta)}
\end{align*}
If we insert $Ms=2^{n(H(W_A)+2\delta)}$ (from (\ref{eq:Msw} later)
and $s=2^{n(I(W_A;RB)+3\delta)}$ (\ref{eq:sw})
we get
\begin{align*}
	T_2 &\leq \frac 1{t(1-\epsilon)}
	2^{-n(I(W_A;B|W_B)-2\delta)}2^{n(I(W;RB)+3\delta)} \nonumber\\
	&\times 2^{-n(I(W_A;W_B)-\delta)}
\end{align*}
Thus we can use
\begin{align*}
  t=2^{-n(I(W_A;B|W_B)-2\delta)}2^{n(I(W;RB)+4\delta)}
	2^{-n(I(W_A;W_B)-\delta)}
\end{align*}

\subsubsection{The measurement faithfully simulates
      $\{\Gamma_z^{AB}\}_z$}\label{sec:faithful}
      
We need to simulate
\begin{align*}
  \Lambda_{z^n}^{AB}=\sum_{u^n,v^n:g^n(u^n,v^n)=z^n}\Lambda^A_{u^n}
  \otimes \Lambda^B_{v^n}
\end{align*}
An alternative is as follows
\begin{align*}
  \Lambda_{z^n}'&=\sum_{w_A^n,w_B^n:\tilde g^n(w_A^n,w_B^n)=z^n}\Lambda^A_{w^n} \otimes \Lambda^B_{w^n} \nonumber\\
  &= \sum_{w_A^n,w_B^n:\tilde g^n(w_A^n,w_B^n)=z^n}\nonumber\\
  &\sum_{u^n,v^n:p(w_A^n,w_B^n|u^n,v^n)>0} p(w_A^n,w_B^n|u^n,v^n)
  \Lambda^A_{u^n}\otimes \Lambda^B_{v^n}\nonumber\\
  &= \sum_{u^n,v^n:g^n(u^n,v^n)=z^n}\nonumber\\
  &\left(\sum_{w_A^n,w_B^n:\tilde g^n(w_A^n,w_B^n)=z^n}p(w_A^n,w_B^n|u^n,v^n)\right)
  \Lambda^A_{u^n}\otimes \Lambda^B_{v^n} \nonumber\\
  &= \Lambda_{z^n}^{AB}
\end{align*}
Thus, we can equivalently prove simulation of $\Lambda_{z^n}'$

From the previous step we know that Bob can decode
$w_A^n(j,m)$ with high probability.
Let $\mathcal{S}_z=\{z^n:\exists j,m,w^n_B:\tilde g^n(w^n_A(j,m),w^n_B)=z^n\}$.
For $z^n\in\mathcal{S}_z$ we consider the following collection of operators
\begin{align*}
  \tilde\Lambda^{AB}_{z^n}
  &=\frac 1{M}\sum_m \sum_{w_B^n,j:\tilde g^n (w_A^n(j,m),w_B^n)=z^n}
  \Gamma^{(m)}_j\otimes \Lambda_{w^n_B}^B \nonumber\\
  &= \sum_{w_A^n,w_B^n:\tilde g^n(w_A^n,w_B^n)=z^n}\frac {c(w_A^n)}{Ms}\frac{S}{1+\epsilon}\nonumber\\&\times\sqrt{\omega^A}^{-1}\xi_{w_A^n}\sqrt{\omega^A}^{-1}\otimes \Lambda_{w^n_B}^B \nonumber\\
  &= \sum_{w_A^n,w_B^n:\tilde g^n(w_A^n,w_B^n)=z^n}\tilde\Lambda_{w^n_A}\otimes \Lambda_{w^n_B}^B
\end{align*}
where
$
c(w_A^n)=\left|\{m,j:w_A^n(m,j)=w_A^n\}\right|
$
We need to evaluate
\begin{align}
  d &= \sum_{z^n}\left\|\sqrt{\omega}
  (\Lambda_{z^n}'-\tilde\Lambda_{z^n}^{AB})\sqrt{\omega}
  \right\|_1 \nonumber\\
  &= \sum_{z^n\notin \mathcal{S}_z}\left\|\sqrt{\omega}
  \Lambda_{z^n}'\sqrt{\omega}
  \right\|_1+\sum_{z^n\in \mathcal{S}_z}\left\|\sqrt{\omega}
  (\Lambda_{z^n}'-\tilde\Lambda_{z^n}^{AB})\sqrt{\omega}
  \right\|_1\nonumber\\
  &\leq \epsilon+\sum_{z^n\in \mathcal{S}_z}
  \left\|\sqrt{\omega}\!\!\!\!\!\!\sum_{w_A^n,w_B^n:\tilde g^n(w_A^n,w_B^n)=z^n}
  \!\!\!\!\!\!\!\!\!\!\!\!(\tilde\Lambda_{w^n_A}
  -\Lambda^A_{w^n_A})\otimes \Lambda_{w^n_B}^B\sqrt\omega
  \right\|_1 \nonumber\\
  &\leq \epsilon+\sum_{w_A^n\notin T_\delta^{W_A^n},w_B^n}
  \left\|\sqrt{\omega}
  (\Lambda^A_{w^n_A}\otimes \Lambda_{w^n_B}^B)\sqrt\omega
  \right\|_1 \nonumber\\
  &+\sum_{w_A^n\in T_\delta^{W_A^n},w_B^n}
  \left\|\sqrt{\omega}
  (\tilde\Lambda_{w^n_A}
  -\Lambda^A_{w^n_A})\otimes \Lambda_{w^n_B}^B\sqrt\omega
  \right\|_1 \label{eq:wawbbound}\\
  &\leq \epsilon+\sum_{w_A^n\notin T_\delta^{W_A^n}}
  \left\|\sqrt{\omega^A}
  \Lambda^A_{w^n_A}\sqrt{\omega^A}
  \right\|_1 \nonumber\\
  &+\sum_{w_A^n\in T_\delta^{W_A^n}}
  \left\|\sqrt{\omega^A}
  (\tilde\Lambda_{w^n_A}
  -\Lambda^A_{w^n_A})\sqrt{\omega^A}
  \right\|_1 \nonumber\\
  &\leq 2\epsilon 
  +\sum_{w_A^n\in T_\delta^{W_A^n}}
  \left\|\sqrt{\omega^A}
  (\tilde\Lambda_{w^n_A}
  -\Lambda^A_{w^n_A})\sqrt{\omega^A}
  \right\|_1 \nonumber\\
  &= 2\epsilon+\sum_{w_A^n\in T_\delta^{W_A^n}}
  \left\|\frac S{1+\epsilon}
  \frac{c(w_A^n)}{Ms}\xi_{w^n_A}
  -p_A(w_A^n)\hat\rho_{w_A^n}
  \right\|_1 \label{eq:dbound} 
\end{align}
The first inequality follows from $P(\mathcal{S}_z)\geq 1-\epsilon$ due to classical typicality, the second inequality from
the triangle inequality and reorganizing the sums,
the third inequality from classical typicality
and \cite[Lemma 3]{AtifHeidariSandeep22}.

We continue to bound the second term
\begin{align}
  d_2  &\leq \sum_{w_A^n\in T_\delta^{W_A^n}}
  p_A(w_A^n)\left\|\xi_{w^n_A}
  -\hat\rho_{w_A^n}
  \right\|_1\nonumber\\
  &+\sum_{w_A^n\in T_\delta^{W_A^n}}
  \left\|\frac S{1+\epsilon}
  \frac{c(w_A^n)}{Ms}\xi_{w^n_A}
  -p_A(w_A^n)\xi_{w_A^n}
  \right\|_1 \nonumber\\
  &\leq \sum_{w_A^n\in T_\delta^{W_A^n}}
  p_A(w_A^n)\left\|\xi_{w^n_A}
  -\xi'_{w_A^n}
  \right\|_1\nonumber\\
  &+\sum_{w_A^n\in T_\delta^{W_A^n}}
  p_A(w_A^n)\left\|\xi'_{w^n_A}
  -\hat\rho_{w_A^n}
  \right\|_1 \nonumber\\
  &+\sum_{w_A^n\in T_\delta^{W_A^n}}
  \left|\frac 1{1+\epsilon}
  \frac{c(w_A^n)}{Ms}
  -\frac 1 S p_A(w_A^n)
  \right| \label{eq:d2w}
\end{align}
The first two terms can be shown to be less
than some $\epsilon''$ exactly as in the proof
of Theorem \ref{thm:Ubound}.
We  bound the third term in (\ref{eq:d2w}). We use the operator Chernoff bound
\cite[Lemma 17.3.1]{WildeBook}. Let $P$ be
the diagonal matrix with $\tilde p_A^n(w_A^n)$ on
the diagonal for all $w_A^n\in T_\delta^{W_A^n}$, and let
$C$ be the same for  the empirical frequencies $\frac{c(w_A^n)}{Ms}$. We have $E[C]=P$ and
$P\geq 2^{-n(H(W_A)+\delta)}I$. Let $E_0$ be the event
that
\begin{align*}
  (1-\epsilon)P\leq C\leq (1+\epsilon)P
\end{align*}
The operator Chernoff bound then gives
\begin{align*}
  P(E_0^c) &\leq 2\cdot 2^{-n(H(W_A)+\delta)}
  \exp\left(-\frac{Ms\epsilon 2^{-n(H(W_A)+\delta)}}{4\ln 2}\right)
\end{align*}
Thus, if 
\begin{align}
Ms\geq 2^{-n(H(W_A)+2\delta)}, \label{eq:Msw}
  \end{align}
this probability
converges to zero. We can then bound the third term in (\ref{eq:d2w})
conditioned on $E_0^c$:
\begin{align*}
\MoveEqLeft
  \sum_{w_A^n\in T_\delta^{W_A^n}}
  \left|\frac 1{1+\epsilon}
  \frac{c(w_A^n)}{Ms}
  -\frac 1 S p_A^n(w_A^n)
  \right| = \left\|\frac 1{1+\epsilon}C-P\right\|_1 \nonumber\\
  &\leq \frac 1{1+\epsilon}(\|\epsilon P\|_1+\|C-P\|_1) \leq \frac{2\epsilon}{1+\epsilon}
\end{align*}

Finally, we will show that the second term in
(\ref{eq.Pe2}) is less than $\epsilon$.
Define
\begin{align*}
  \MoveEqLeft\mathcal{M}_{\Lambda_{w^n_A}^A\otimes \Lambda_{w^n_B}^B}(\phi) \nonumber\\
  &=\sum_{w_A^n,w_B^n}\Tr_{A^n}\left\{\left(\Lambda_{w^n_A}^A\otimes \Lambda_{w^n_B}^B\right)\phi\right\}\otimes|w_A^n\rangle\langle w_A^n|
  \otimes |w_B^n\rangle\langle w_B^n|
\end{align*}
and similar for $\tilde\  $. The following lemma
is a slight generalization of \cite[Lemma 4]{WildeAl2012}
\begin{lem}
\begin{align}
\MoveEqLeft\sum_{w_A^n,w_B^n}
  \left\|\sqrt{\omega}
  (\tilde\Lambda_{w^n_A}^A
  -\Lambda^A_{w^n_A})\otimes \Lambda_{w^n_B}^B\sqrt\omega
  \right\|_1 \label{eq:faithful2a}\\
  &=\left\|\left(I^{R^n}\otimes \mathcal{M}_{\tilde\Lambda_{w^n_A}^A\otimes \Lambda_{w^n_B}^B}\right)(\phi^{RAB})\right.\nonumber\\
  &\left.-\left(I^{R^n}\otimes \mathcal{M}_{\Lambda_{w^n_A}^A\otimes \Lambda_{w^n_B}^B}\right)(\phi^{RAB})\right\|_1 \end{align}
	
\end{lem}
The modification to the proof
of \cite[Lemma 4]{WildeAl2012} is
to replace the reference system
$R$ with $RB$, which here purifies
$A$, and replace $\Lambda_x^A$ with
$\Lambda_{w_A}^A\otimes \Lambda_{w_B}^B$
and the proof will then be identical.

Now
\begin{align}
\MoveEqLeft\left\|\left(I^{R^n}\otimes \mathcal{M}_{\Lambda_{w^n_A}^A\otimes \Lambda_{w^n_B}^B}\right)(\phi^{RAB})\right.\nonumber\\
  &\left.-\left(I^{R^n}\otimes \mathcal{M}_{\Lambda_{w^n_A}^A\otimes \Lambda_{w^n_B}^B}\right)(\phi^{RAB})\right\|_1 \nonumber\\
  &\geq \left\|\Tr_{R^n}\left\{\left(I^{R^n}\otimes \mathcal{M}_{\tilde\Lambda_{w^n_A}\otimes \Lambda_{w^n_B}^B}\right)(\phi^{RAB})\right\}\right.\nonumber\\
  &\left.-\Tr_{R^n}\left\{\left(I^{R^n}\otimes \mathcal{M}_{\Lambda_{w^n_A}^A\otimes \Lambda_{w^n_B}^B}\right)(\phi^{RAB})\right\}\right\|_1
  \label{eq:faithfu2b}
\end{align}
Here
\begin{align*}
  \MoveEqLeft
  \Tr_{R^n}\left\{\left(I^{R^n}\otimes \mathcal{M}_{\Lambda_{w^n_A}^A\otimes \Lambda_{w^n_B}^B}\right)(\phi^{RAB})\right\} \nonumber\\
  &=\sum_{w_A^n,w_B^n}\Tr_{A^n}\left\{\Tr_{R^n}\left(I^{R^n}\otimes\Lambda_{w^n_A}^A\otimes \Lambda_{w^n_B}^B\right)(\phi^{RAB})\right\}\nonumber\\&\otimes|w_A^n\rangle\langle w_A^n|
  \otimes |w_B^n\rangle\langle w_B^n|\nonumber\\
  &=\sum_{w_A^n,w_B^n}\Tr_{A^n}\left\{(\Lambda^A_{w_A^n}\otimes \Lambda_{w_B^n}^B)(\rho^{AB})^{\otimes n}\right\}\nonumber\\&\otimes|w_A^n\rangle\langle w_A^n|   \otimes |w_B^n\rangle\langle w_B^n|\nonumber\\
  &=\sum_{w_A^n,w_B^n}p(w_A^n,w_B^n)\rho_{w_A^n,w_B^n}^{B^n}\otimes|w_A^n\rangle\langle w_A^n|
  \otimes |w_B^n\rangle\langle w_B^n|
\end{align*}
Similarly
\begin{align*}
  \MoveEqLeft
  \Tr_{R^n}\left\{\left(I^{R^n}\otimes \mathcal{M}_{\tilde \Lambda_{w^n_A}^A\otimes \Lambda_{w^n_B}^B}\right)(\phi^{RAB})\right\} \nonumber\\
  &=\!\!\!\!\sum_{i,k,m:k\in\mathcal{S}^{(m)}(i,w_B^n)}\!\!\!\!\!\!\!\!\!\!\!\!\!\!\!\!p(i,k,m,w_B^n)I_{w_A^n=w_A^n(i,k,m)}
  \tilde\rho^{B^n}_{w_B^n,w_A^n(i,k,m)}
\end{align*}
Then 
\begin{align}
\MoveEqLeft  \left\|\Tr_{R^n}\left\{\left(I^{R^n}\otimes \mathcal{M}_{\tilde\Lambda_{w^n_A}\otimes \Lambda_{w^n_B}^B}\right)(\phi^{RAB})\right\}\right.\nonumber\\
  &\left.-\Tr_{R^n}\left\{\left(I^{R^n}\otimes \mathcal{M}_{\Lambda_{w^n_A}^A\otimes \Lambda_{w^n_B}^B}\right)(\phi^{RAB})\right\}\right\|_1 \nonumber\\
  &=\sum_{w_A^n,w_B^n}\left\|\vphantom{\sum_{i,k,m}}p(w_A^n,w_B^n)\rho^{B^n}_{w_A^n,w_B^n}\right.\nonumber\\
  &\left.-\!\!\!\!\sum_{i,k,m:k\in\mathcal{S}^{(m)}(i,w_B^n)}\!\!\!\!\!\!\!\!\!\!\!\!\!\!\!\!p(i,k,m,w_B^n)I_{w_A^n=w_A^n(i,k,m)}
  \tilde\rho^{B^n}_{w_B^n,w_A^n(i,k,m)}\right\|_1\nonumber\\
  &\geq\sum_{w_A^n\in T_\delta^{W_A^n},w_B^n\in T_\delta^{W_B^n}}\left\|\vphantom{\sum_{i,k,m}}p(w_A^n,w_B^n)\rho^{B^n}_{w_A^n,w_B^n}\right.\nonumber\\
  &\left.-\!\!\!\!\sum_{i,k,m:k\in\mathcal{S}^{(m)}(i,w_B^n)}\!\!\!\!\!\!\!\!\!\!\!\!\!\!\!\!p(i,k,m,w_B^n)I_{w_A^n=w_A^n(i,k,m)}
  \tilde\rho^{B^n}_{w_B^n,w_A^n(i,k,m)}\right\|_1 \label{eq:bbbound}
\end{align}
Here (\ref{eq:faithful2a})
is bounded by some $\epsilon''$ in
(\ref{eq:wawbbound}) with high probability. On the hand it
also always bounded by 1, and therefore the expectation 
is also bounded by some arbitrarily small    $\epsilon'''$. 
And the expectation of (\ref{eq:bbbound}) is the second term in
(\ref{eq.Pe2}).

All of these bounds were under the assumption that no error
occurs. But if an error occurs, the contribution to
$d$ is at most 1, and the total error probability converges
to zero with $n$. 

\section{Discussion of the Results}
It would be desirable to have a single rate region
that includes the two rate regions in this paper,
not just a time-sharing region. However,
the bound $\log s>I(U;RB)$ (here $\log s$ is the
communication rate prior to binning) in (\ref{eq:scond})
is fundamentally due to the use of conditional typicality with $u^n$
in (\ref{eq:xiprime}) and difficult to come around.
The scheme in Theorem \ref{thm:Ubound} could
be combined with using quantum side information
as in Theorem \ref{thm:wmeasure}. In fact, 
the decoding projections (\ref{eq:Pijkm}) can still
be used. However, then one ends up with the
same closeness of states condition as in (\ref{eq.Pe2}).
Working through the proof, one then sees that
the condition for this is (\ref{eq:Msw}),
and one therefore ends up with the bound (\ref{eq:WRSbound}), whereby Theorem
\ref{thm:Ubound} always would be worse than
Theorem \ref{thm:wmeasure}.

It would also be desirable to have a converse. However,
even in the classical case, the converse is 
tricky. It relates $H_G(U|V)$ to  Wyner-Ziv
rate distortion with side information in the limit
of zero distortion. The Wyner-Ziv rate distortion region
is known in the classical case (as a single letter expression), but in
the quantum case, as far as we know, only as a non-regularized expression \cite{LuoDevetak09}. And,
as the paper \cite{LuoDevetak09} states, the non-regularized converse is
actually trivial. So, it seems not easy to get a
good converse.


\bibliographystyle{IEEEtran}
\bibliography{Coop06,ahmref2,Coop03,BigData,Quantum}

\end{document}